\documentclass[aps,pre,twocolumn,showpacs,floatfix]{revtex4}
\usepackage{graphicx,color}
\usepackage{amsmath,amssymb}
\usepackage{hyperref}

\newcommand{\phI}{Z$_{\rm a}$}
\newcommand{\phII}{Z$_{\rm d}$}
\newcommand{\phIII}{U$_{\rm ad}$}
\newcommand{\phIV}{U$_{\rm dd}$}

\newcommand{\figmodel}{ 
\begin{figure}[tb] 
	\centering
	\includegraphics[height=1.0in]{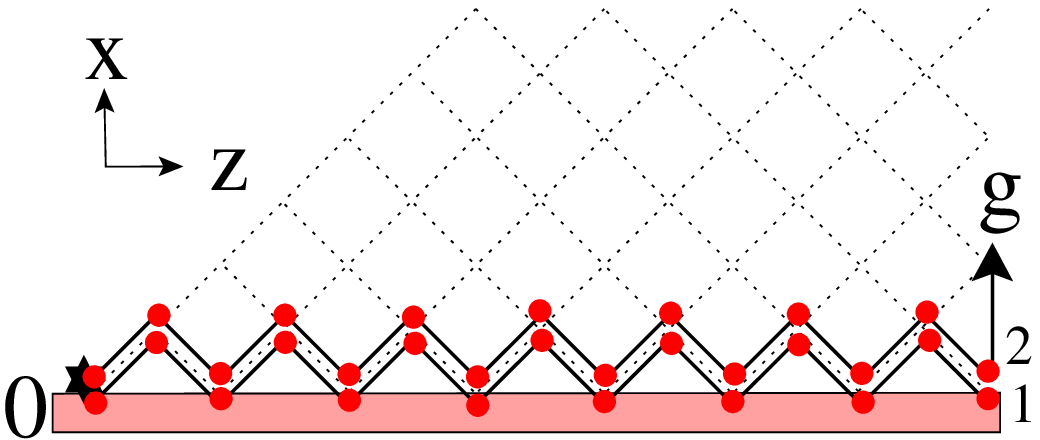} 
	
	\caption{ Schematic diagram of a dsDNA adsorbed on the surface
	(shaded region) in $1+1$ dimensional square lattice. There is a
	binding energy $-\epsilon_{\mathrm{b}}$ between bases (shown by
	filled circles) of the strands of the DNA. One end of the DNA is
	always kept anchored on the surface at the origin (shown by star).
	The free strand (denoted by 1) can gain energy
	$-\epsilon_{\mathrm{w}}$ for every contact with the surface (i.e.,
	$x_1 = 0$). An external force $\mathrm{g}$ (shown by arrow) is
	applied at the free end of the pulled strand (denoted by 2) in the
	transverse direction.
	}\label{fig:model}
		
\end{figure} }

\newcommand{\figphzig}{
\begin{figure}[tb]
	\centering
	\includegraphics[width=2.5in]{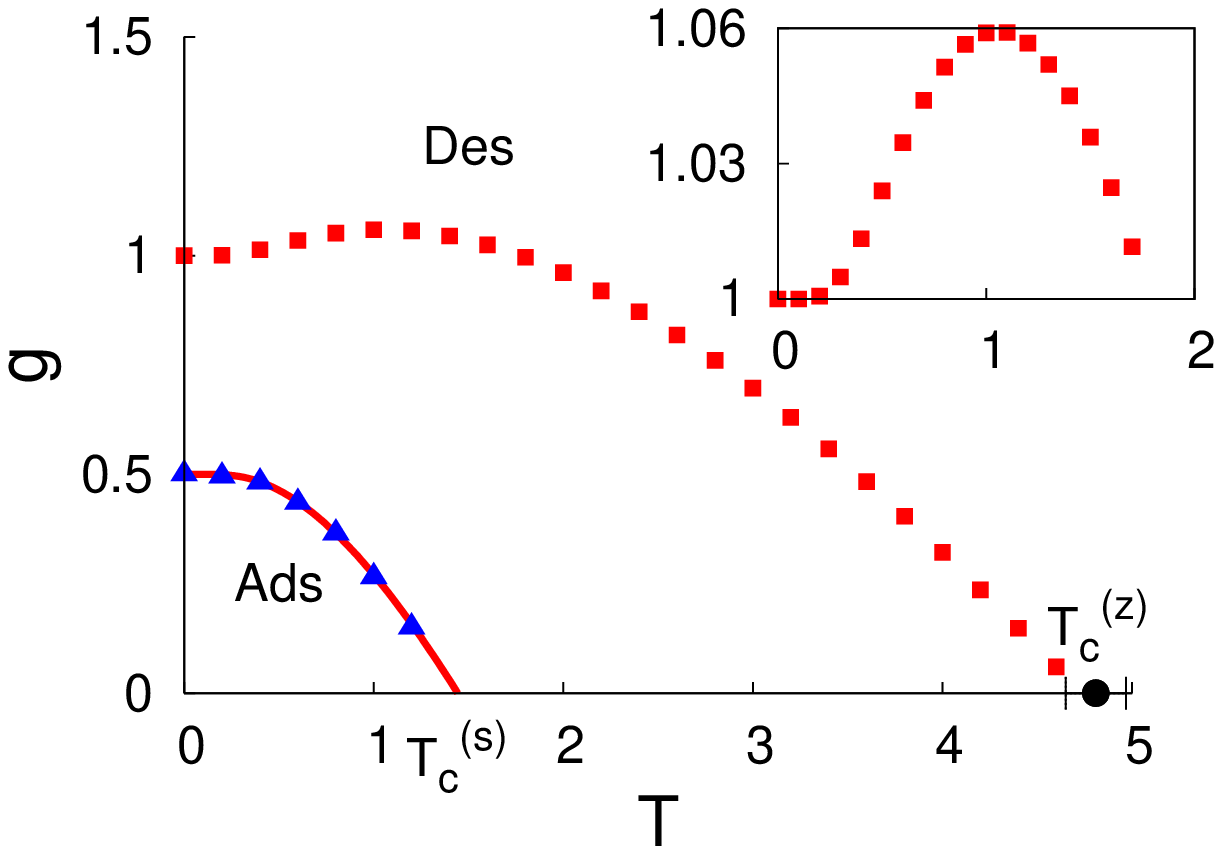}
    
	\caption{ $g$ vs $T$ phase diagram for the unzipping of an adsorbed
	polymer from an attractive impenetrable surface.  The points are
	obtained by using the exact transfer matrix and the solid curve is
	an analytical result (see Eq.~(\ref{eq:gcpoly})). The triangles and
	squares represent the phase boundary between the adsorbed (Ads) and
	the desorbed (Des) phases for the straight and zig zag hard-walls
	respectively. The thermal desorption temperatures for the straight
	and the zig zag surfaces are shown respectively by $T_c^{(\rm s)}$
	and $T_c^{(\rm z)}$. For the later case it is shown by a circle.
	}\label{phzig}

\end{figure}
}

\newcommand{\figxgzig}{
\begin{figure}[tb]
	\centering
	\includegraphics[width=2.5in]{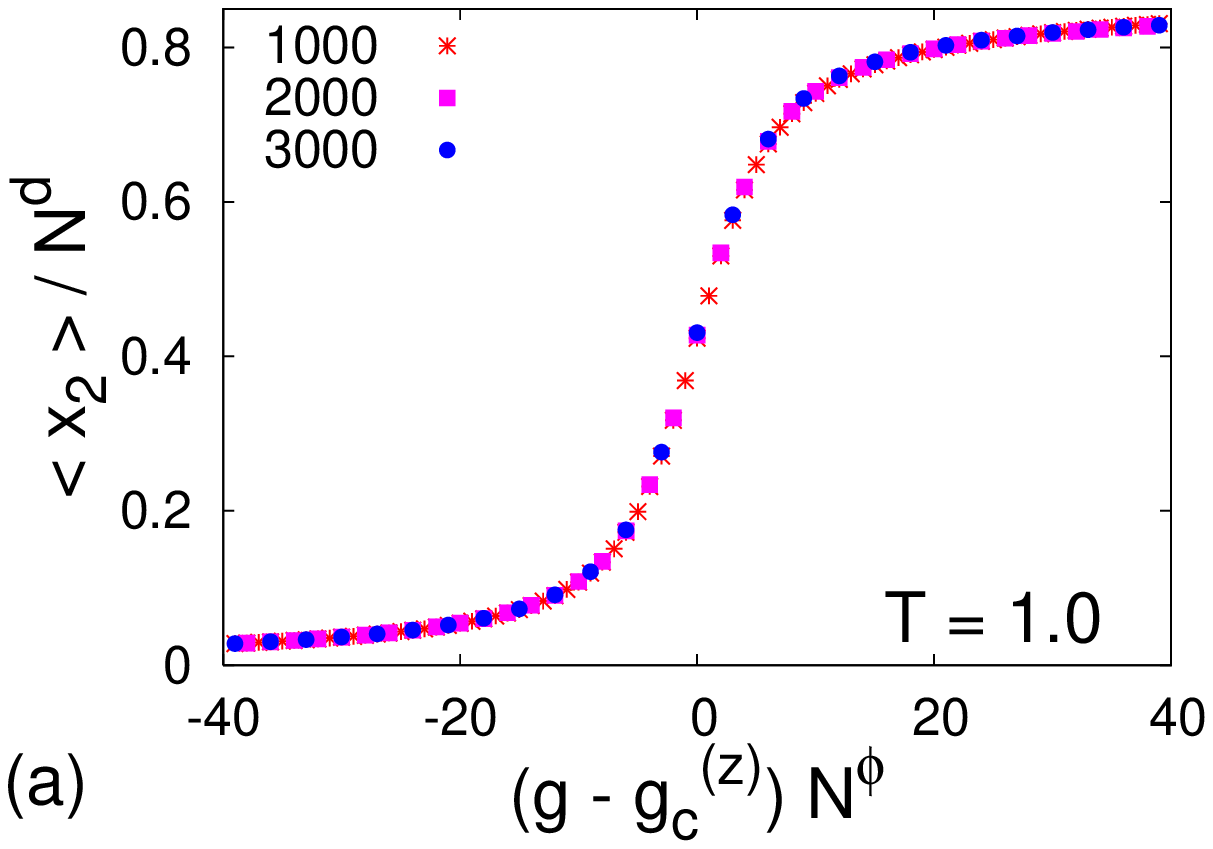}
	\includegraphics[width=2.5in]{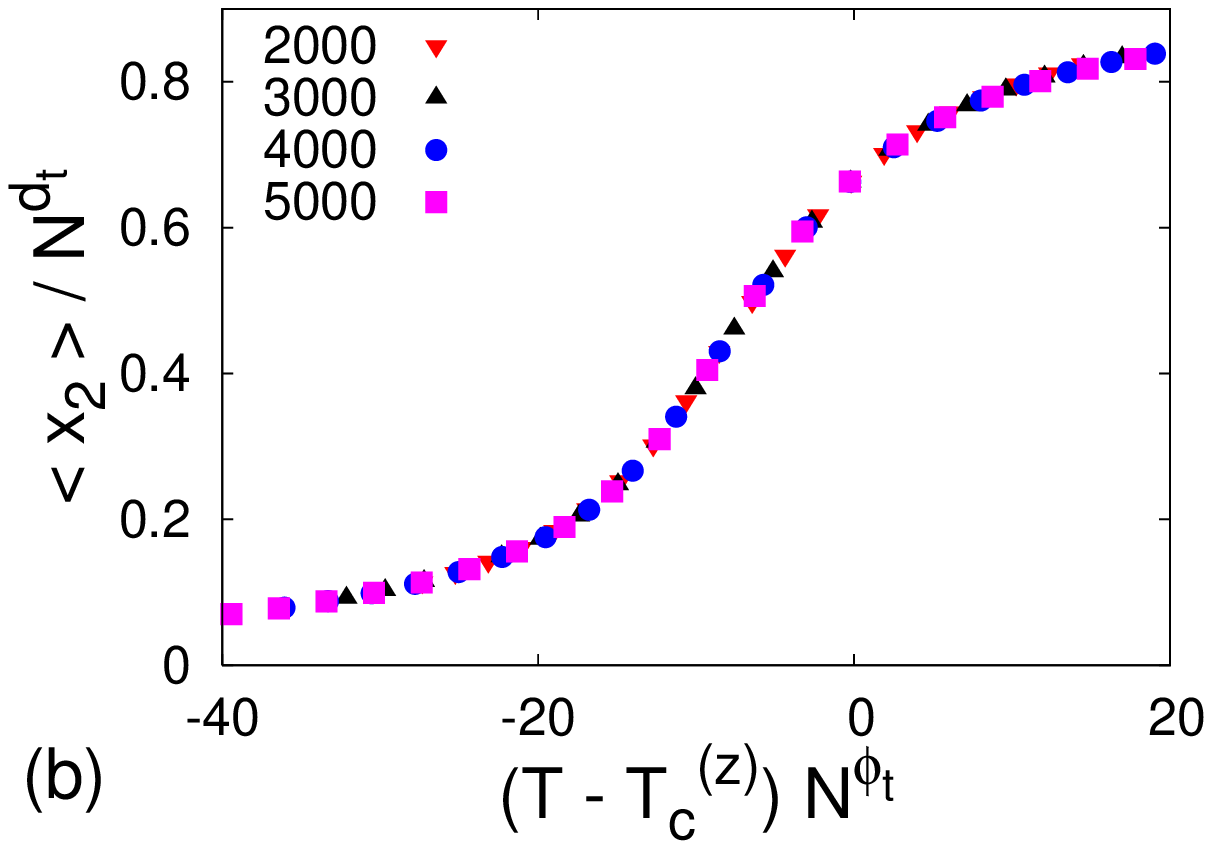}

	\caption{Data collapse of the average distance $\langle x_2 \rangle$
	of the last monomer from the zig zag surface (a) as a function of
	$g$ at $T=1.0$ for $N=1000$, $2000$ and $3000$.  The critical
	exponents are $d = 0.99 \pm 0.01$ and $\phi = 1.0 \pm 0.01$ and the
	critical force $g_{\rm c}^{(z)} = 1.059 \pm 0.001$.  (b) as a
	function of $T$ for $N=2000$, $3000$, $4000$ and $5000$. In this
	case the critical exponents are $d_t = 0.53 \pm 0.03$, $\phi_t = 0.4
	\pm 0.03$ and $T_{\rm c}^{(z)} = 4.81 \pm 0.16$.} \label{xgzig}

\end{figure}
}

\newcommand{\figphdna}{
\begin{figure}[tb]
   \centering
   \includegraphics[width=2.5in]{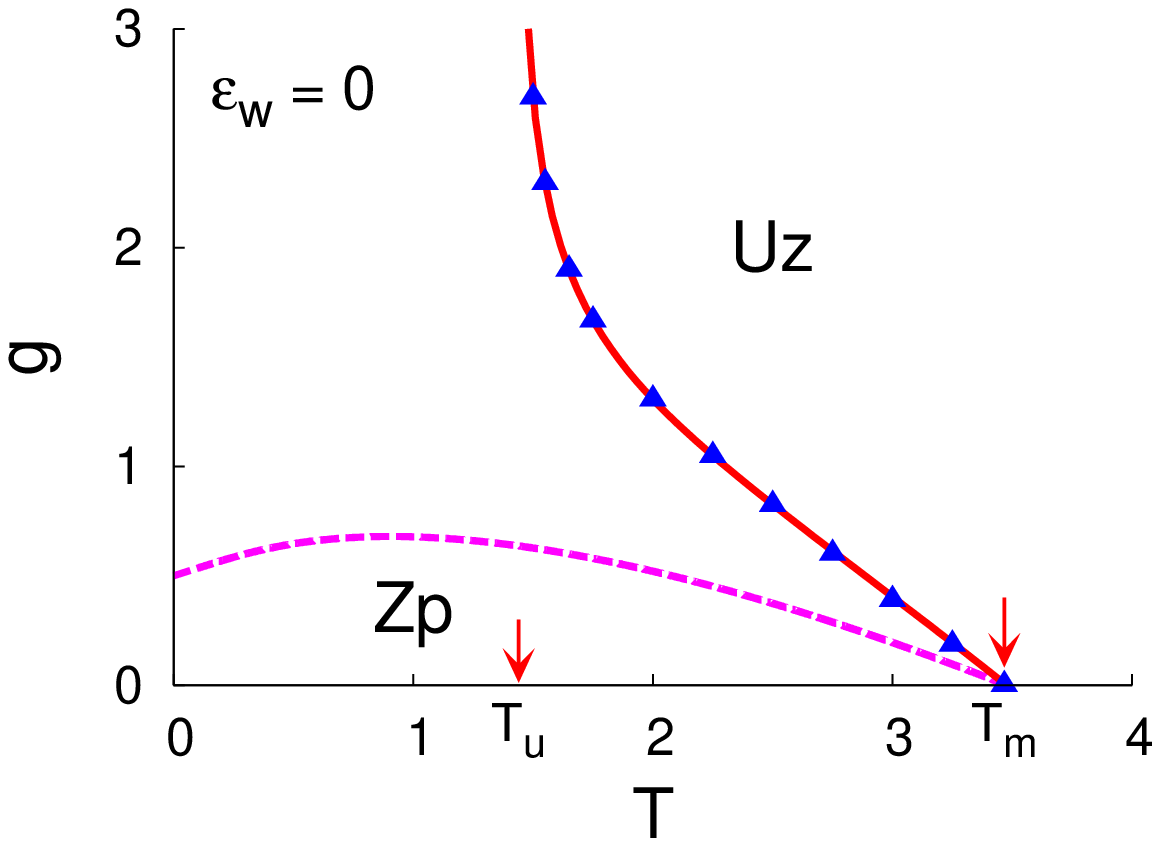}
       
   \caption{$g$ vs $T$ phase diagram for the DNA unzipping. The dashed
   line represents the phase boundary for the unzipping by pulling
   strands in opposite directions by an external force force $g$. The
   solid line is exact phase boundary (Eq.~(\ref{eq:gcsingpull})) and
   the triangles are from the numerics for the pulling of a single
   strand for which the transition takes place only above a minimum
   temperature $T_{\rm u}$.}\label{fig:phdna}

\end{figure}
}

\newcommand{\figxgzero}{
\begin{figure}[b]
	\centering
    \includegraphics[width=2.5in]{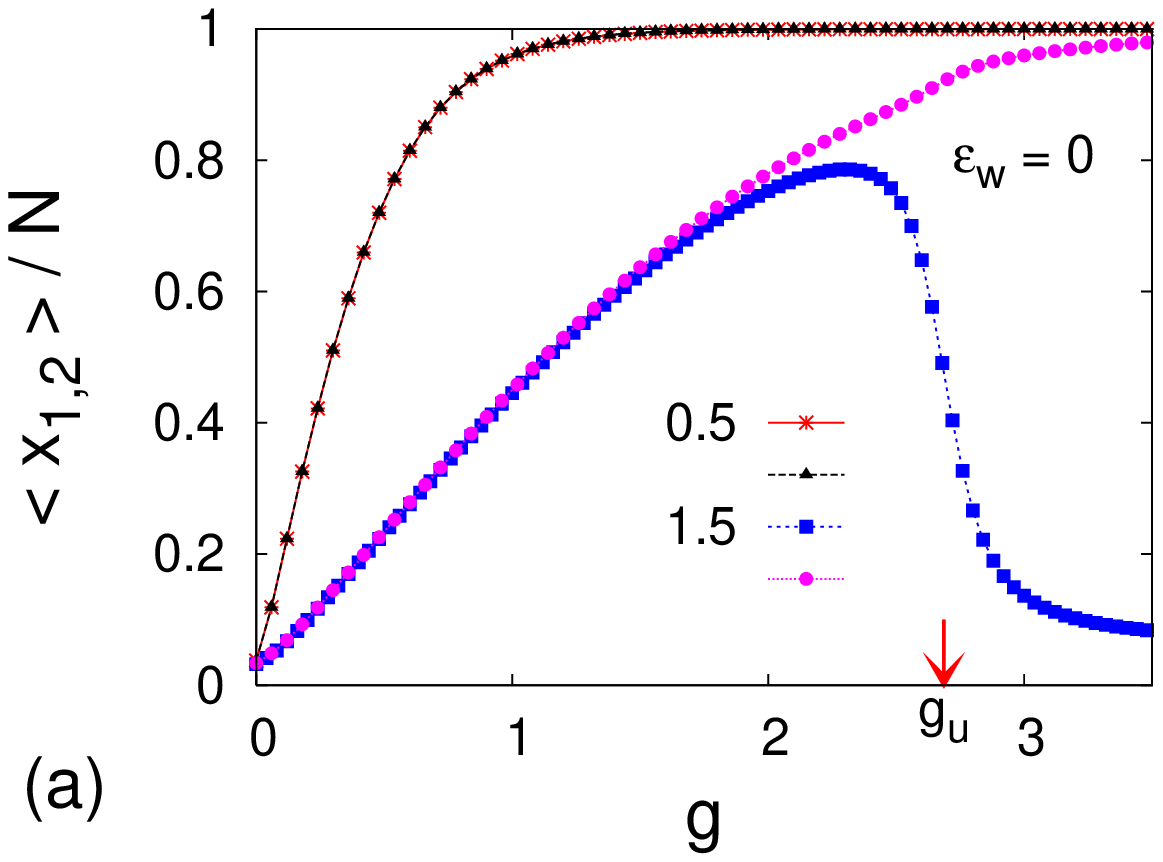}
	\includegraphics[width=2.5in]{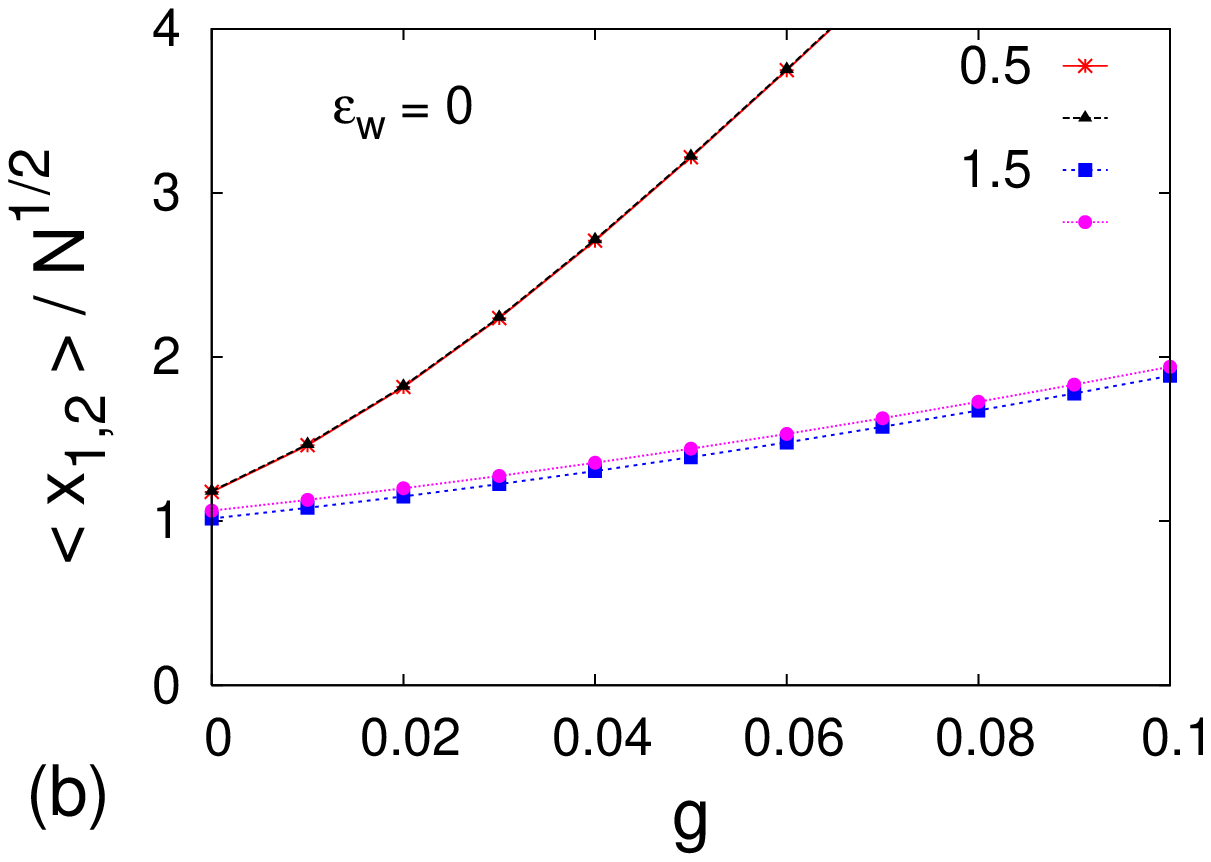}
	
	\caption{(a) ${\rm g}$ vs $\langle x_{1,2} \rangle /N$ isotherms at
	$T = 0.5$ and $1.5$ for $\epsilon_{\rm w} = 0$. The squares and
	circles refer to the two different strands of the dsDNA. The
	critical force above which the dsDNA unzips is shown by $g_{\rm u}$.
	In (b) same isotherms are shown near $g=0$ in a different scale
	$\langle x_{1,2} \rangle / N^{1/2}$. The broken lines are a guide to
	the eyes.  }\label{fig:xg0}

\end{figure}
}

\newcommand{\figextzero}{
\begin{figure}[t]
	\centering
	\includegraphics[width=2.5in]{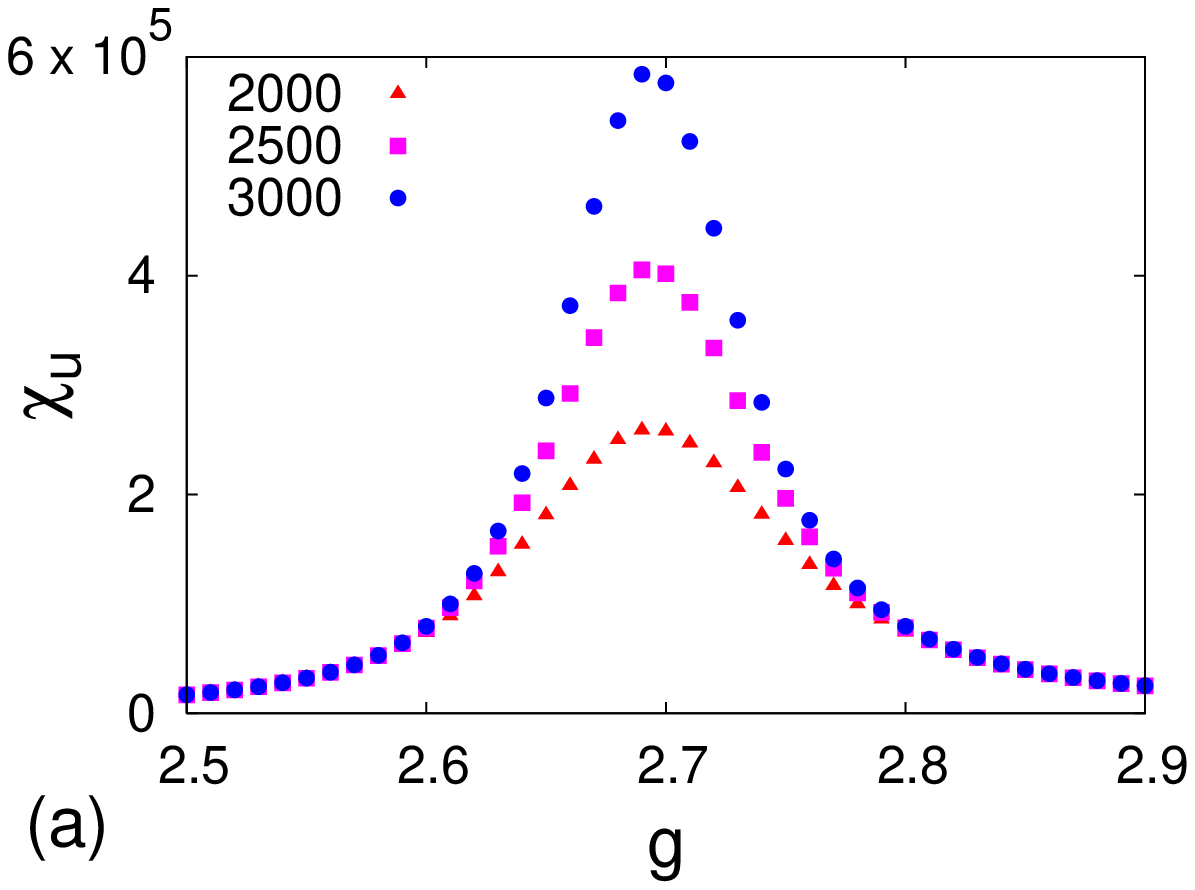}
	\includegraphics[width=2.5in]{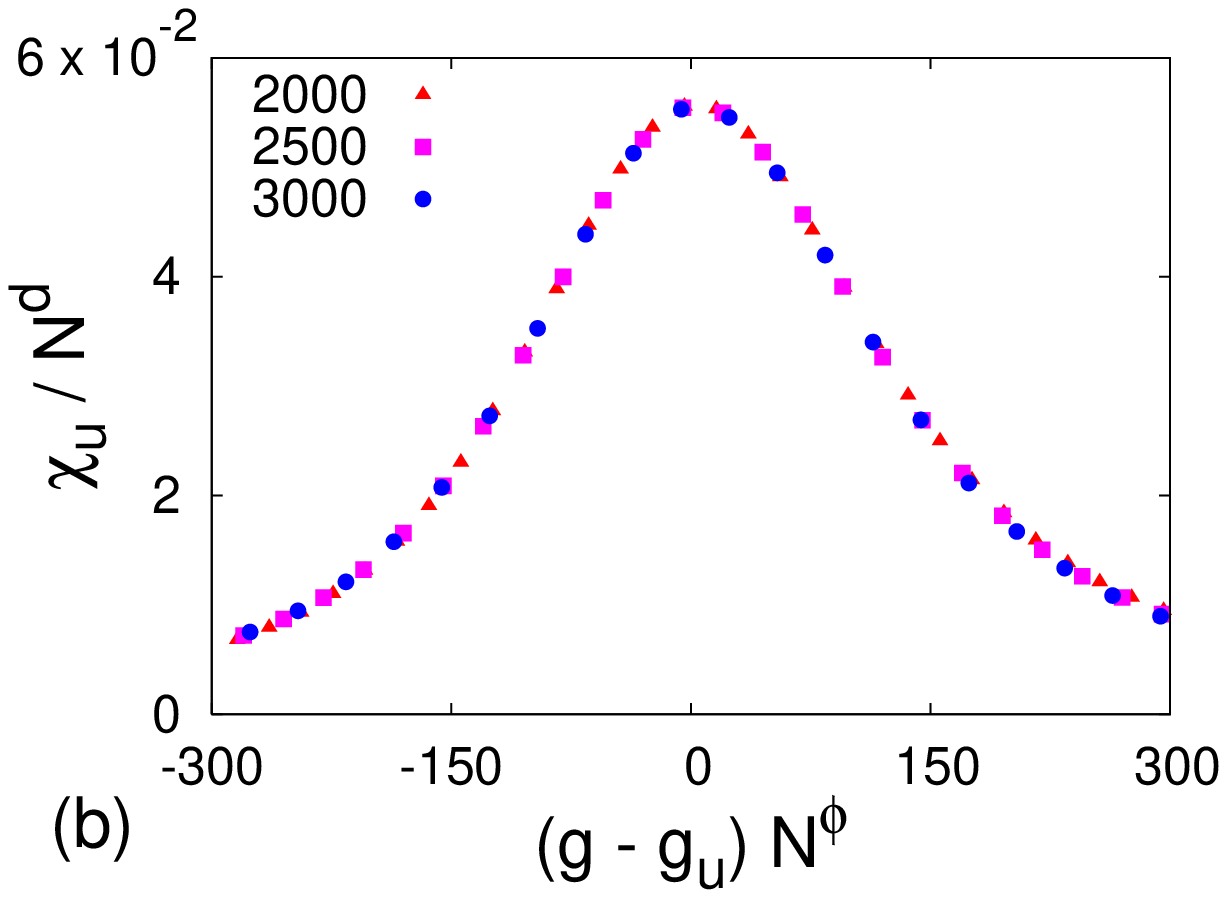}
		
	\caption{(a) Isothermal extensibility $\chi_u$ vs $g$ at $T=1.5$ for
	the DNA of lengths $N=2000$, $2500$ and $3000$.  (b) The data
	collapse of the extensibility. For all plots $\epsilon_{\rm b} = 1$.
	}\label{fig:ext0}

\end{figure}
}

\newcommand{\figxg}{
\begin{figure}[tb]
	\centering
	\includegraphics[width=2.5in]{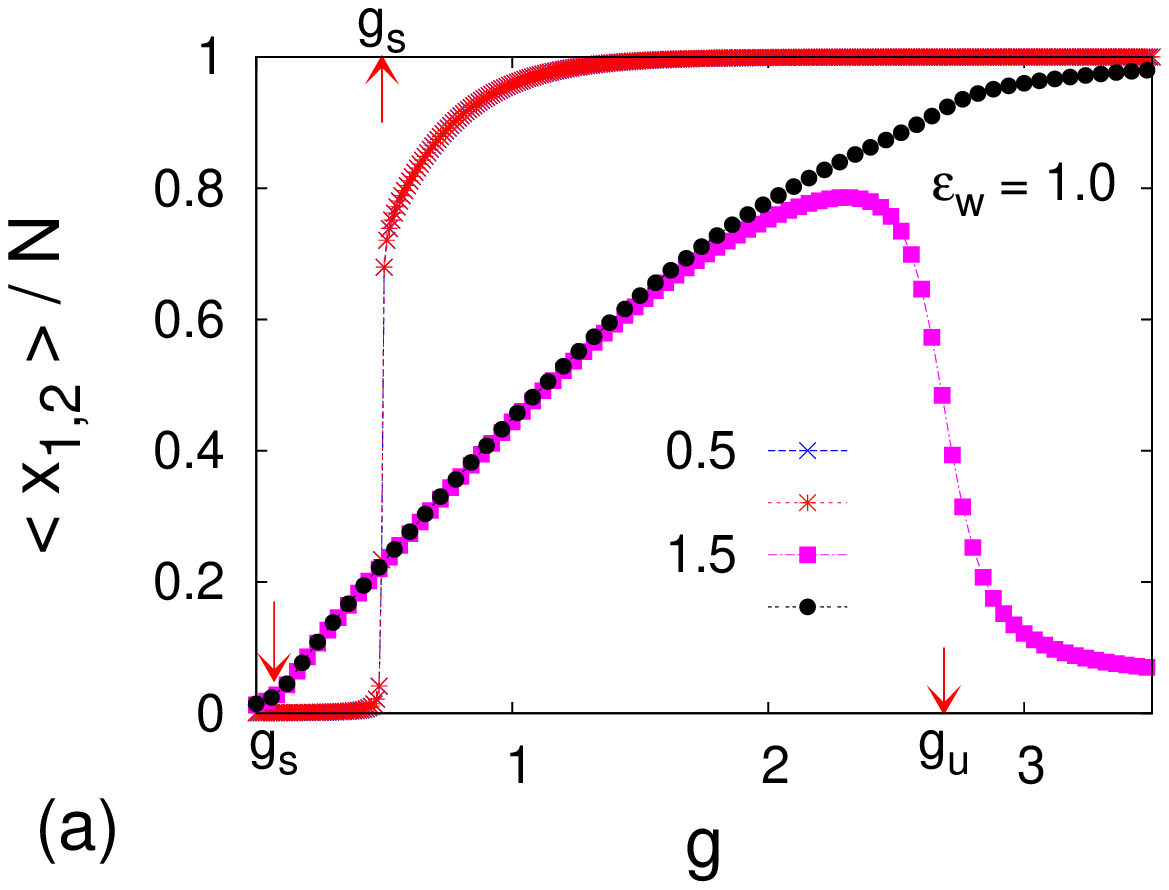}
	\includegraphics[width=2.5in]{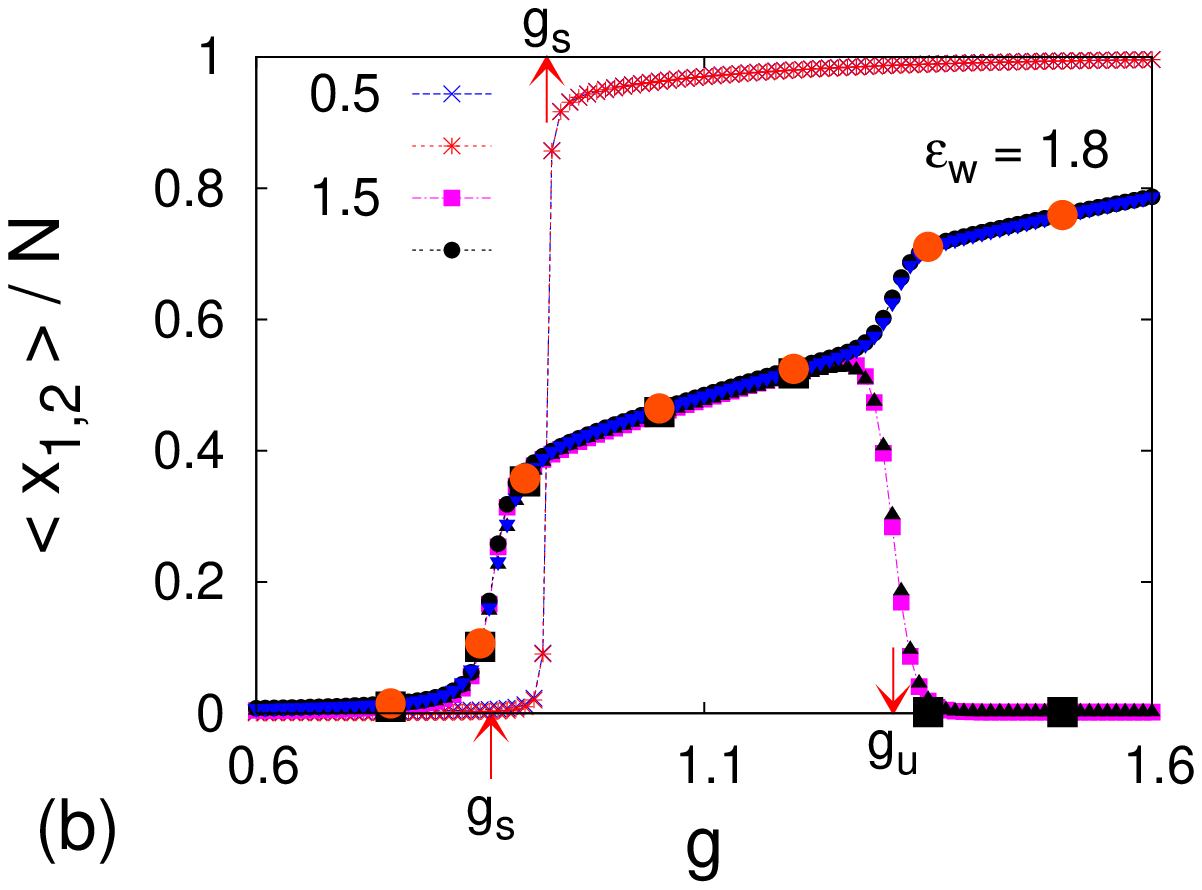}
	\includegraphics[width=2.5in]{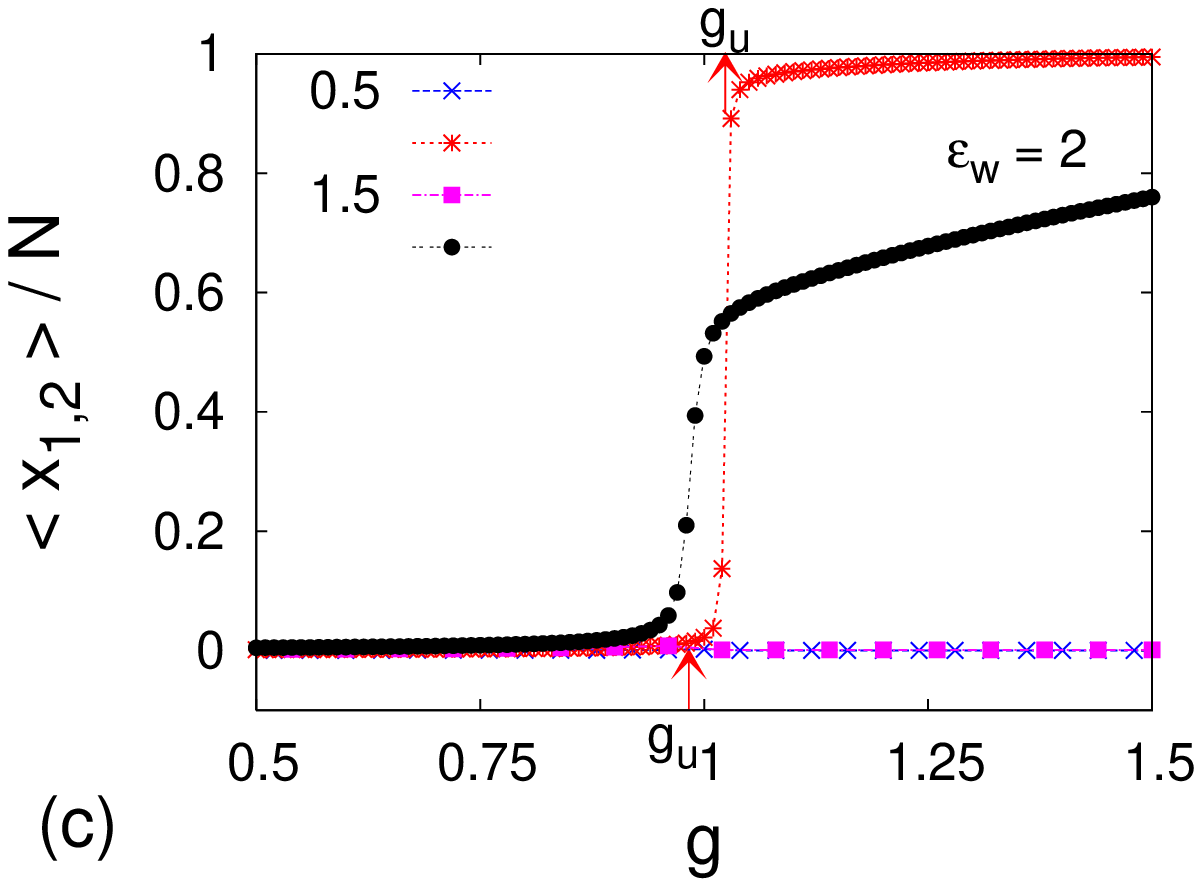}

	\caption{The $\langle x_{1,2} \rangle /N$ vs $g$ isotherms at
	temperatures, $T=0.5$ and $1.5$ for various $\epsilon_{\rm w}$. All
	isotherms are for $N=1000$ with $\epsilon_{\rm b} = 1$. The
	transitions Uz and Sz take place at critical forces $g_u$ and $g_s$
	respectively (see text). These are shown by arrows in the plot. (a)
	For $\epsilon_{\rm w} = 1$, (b) For $\epsilon_{\rm w} = 1.8$. We
	have also shown the averages obtained by using Monte Carlo
	simulations. The squares and circles refer to the two different
	strands of the dsDNA. The upper and the lower triangles are the
	estimates given by the multiple histogram technique at various $g$.
	(c) For $\epsilon_{\rm w} = 2$.  The broken lines are a guide to the
	eyes.}\label{fig:xg} 

\end{figure} 
}

\newcommand{\figphdia}{
\begin{figure}[tb]
	\centering
	\includegraphics[width=2.5in]{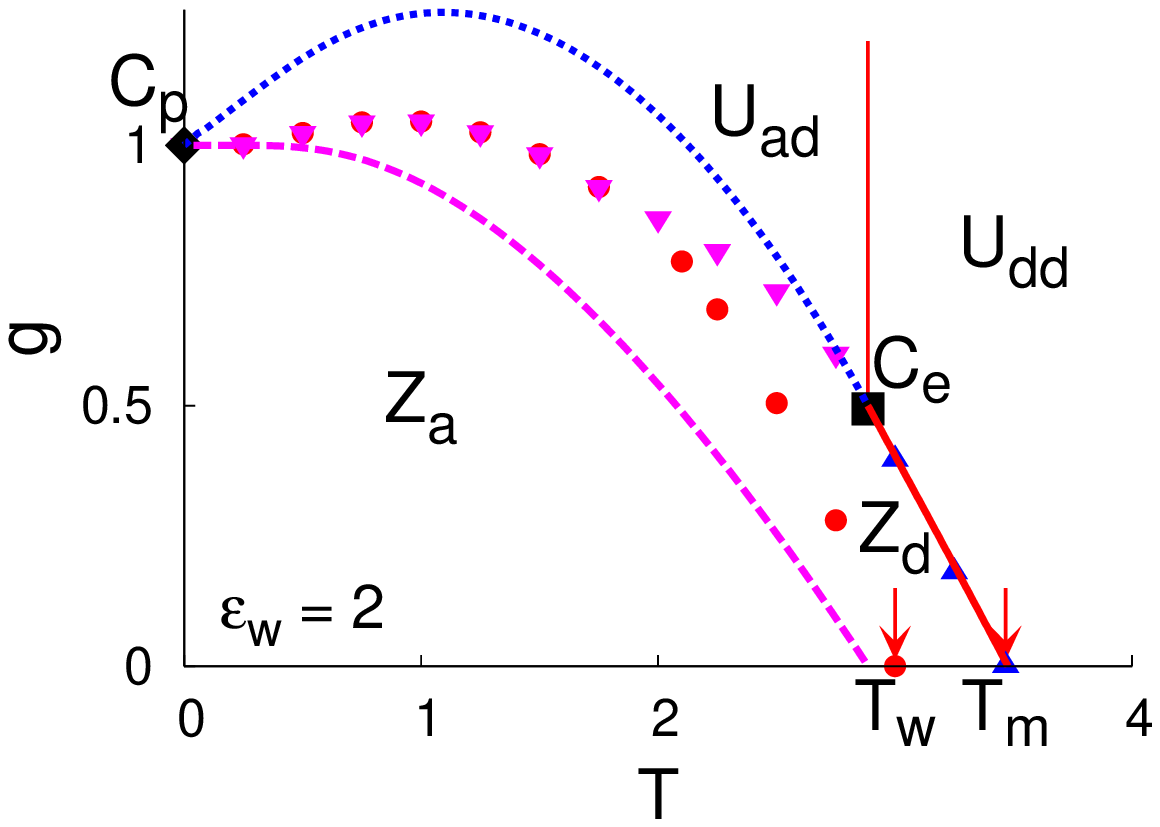}

   \caption{ $g$ vs $T$ phase diagrams for $\epsilon_{\rm w}=2$. The
   points are obtained by using the exact transfer matrix and $T_{\rm
   w}$, and $T_{\mathrm{m}}$ (shown by arrows) represent respectively
   the temperature at which the dsDNA desorbs from the surface, and the
   melting temperature of the DNA. The triple point and the critical end
   point are shown by a diamond and a square and represented
   respectively by C$_{\mathrm{p}}$ and C$_{\mathrm{e}}$.  The thick
   dashed, solid and dotted lines are respectively
   Eqs.~(\ref{eq:gcpoly}), (\ref{eq:recmixed}) and (\ref{eq:gcsurf}) and
   are approximation to the phase boundaries represented by circles, up
   triangles and down triangles.  }\label{fig:phdia} 

\end{figure} 
}

\begin{document}

\title{Can a double stranded DNA be unzipped by pulling a single
strand?: Phases of adsorbed DNA }
\author{Rajeev Kapri }
\email{rkapri@theory.tifr.res.in}
\affiliation{Department of Theoretical Physics, Tata Institute of
Fundamental Research, 1 Homi Bhabha Road, Mumbai -- 400 005 India.}
\date{\today}
\begin{abstract}
	We study the unzipping of a double stranded DNA (dsDNA) by applying
	an external force on a single strand while leaving the other strand
	free. We find that the dsDNA can be unzipped to two single strands
	if the external force exceeds a critical value. We obtain the phase
	diagram which is found to be different from the phase diagram of
	unzipping by pulling both the strands in opposite directions. In the
	presence of an attractive surface near DNA, the phase diagram gets
	modified drastically and shows richer surprises including a critical
	end point and a triple point.  
\end{abstract}
\maketitle

\section{Introduction}

DNA replication in prokaryotes gets initiated by unzipping of a few base
pairs at one end of the dsDNA and then continues till end. These
processes are assisted by various enzymes, often by exerting force on
DNA~\cite{watson}. It is now known theoretically, that a dsDNA undergoes
an unzipping transition under the action of an external force if the
force exceeds a temperature dependent critical
value~\cite{somen:unzip,sebastian,maren,KapriSMB}. Many studies of this
unzipping
transition~\cite{scaling,kafri,tkachenko,Lubensky,KapriSMB:JPCM} have
revealed the importance of ensembles namely, the fixed distance and the
fixed force ensemble~\cite{smb02}. For the fixed distance case, the
distance between the strands is kept constant and the force required to
maintain the distance is allowed to fluctuate, whose average is the
quantity of interest, whereas, the average of the fluctuating distance
between the strands where an external force is acting is the quantity of
interest in the fixed force ensemble. For the single molecule studies,
the results are known to depend on the ensemble
used~\cite{KapriSMB,Keller}. The helicases work in both the fixed
distance and the fixed force ensembles~\cite{smb04}.

In recent years, the unzipping transition has been studied in detail
with various extensions of the basic model. These include studies of
models with intermediate
phases~\cite{KapriSMB,KapriSMB:JPCM,KumGirSMB,Navin}, dependence on
pulling directions~\cite{KumarGiriPRL}, models with additional features
like semiflexibility~\cite{KierfeldPRL},
heterogeneity~\cite{Lubensky,Chen,Zhou,Allahverdyan,Lam}, saturation of
hydrogen bonding~\cite{GiriKumarPRE}, random forcing~\cite{Kapri:prl07}
etc.  A similar problem of unzipping of an adsorbed polymer from the
surface has also been studied~\cite{kapri:pre05,orlandini}. Experimental
studies use various micro-manipulation
techniques~\cite{Roulet,Anselmetti,Danilowicz}.  In all of these
studies, the focus was on the breaking of the pairing by the external
force. A more complicated situation emerges in DNA replication and
segregation (e.g. in {\it Bacillus subtilis, E. Coli} etc.), where the
membrane-DNA complexes play an important role~\cite{firshein,herrick}.
For example, in Bacillus subtilis, the DnaB protein has been shown to be
a membrane-associated protein that is involved in initiation of
replication~\cite{hoshino87}. Also, it is known that the interaction
between the DNA and the membrane can form ordered domains~\cite{Dan}.
Analogous laboratory situation would involve a substrate-DNA
interaction.  Recently atomic force microscopy has been used to identify
the binding mechanism between the DNA and the DNA-binding agents by
pulling one end of the dsDNA, which is immobilized on a gold surface in
the presence of DNA-binding agents~\cite{eckel}.

In this paper our aim is to study a simple model of unzipping of a dsDNA
by applying an external force on a single strand in the transverse
direction. Can a dsDNA be unzipped? Within the same framework, we also
study the behaviour of the dsDNA in the presence of an attractive
surface near it. This surface can mimic a membrane on which the DNA can
be immobilized. Since adsorption of a polymer (here a dsDNA or a single
strand of dsDNA) is a phase transition by itself (often critical), the
competition between adsorption and unzipping is expected to add new
features to the unzipping phase diagram.

The presence of the additional DNA substrate interaction both {\it
in-vivo} and {\it in-vitro} system requires a different extension of the
hitherto used models for the unzipping transition. A helical structure
could pose extra problems for surface adsorption. In order to focus on
the competition between adsorption and pairing, we use an extension of
the Poland-Scheraga (PS) model~\cite{Poland66} in two dimensions.
Previous studies of DNA unzipping showed that the lattice model
preserves, even in two dimensions, the basic results of DNA unzipping
including the first order nature of the phase transition and the
existence of a reentrant region~\cite{maren}.  For the problem at hand,
we have also done Monte Carlo simulations in 2+1 dimensions, though
without considering the helical structure of DNA, but the strands can
wind together because of the dynamics, and find that the force-distance
isotherms are qualitatively similar to the isotherms obtained in 1+1
dimensions.

At this point it is worthwhile to mention a few caveats of our approach.
To achieve simplicity and tractability, we ignored the helical structure
of dsDNA. The main obstacle in incorporating it is the absence as yet of
any analytically tractable models for DNA melting that admits a helical
ground state. In this situation, a PS type model serves as the starting
point. We will see that if the base pairing energy of the DNA is greater
than the binding to the surface, the unzipping of dsDNA by pulling a
single strand takes place when it is away from the surface.  Since the
other strand of the DNA is left free, it can unwind itself from the
pulled strand and the helical structure of the DNA can be safely
neglected. However, in the opposite limit, the helical structure of the
DNA becomes more and more crucial as the strength of binding to the
surface is increased and its role can no longer be neglected.  Another
feature that may play an important role is the difference in the length
per base pair of dsDNA and single-stranded DNA (ssDNA).  Since it is
difficult to incorporate this in a PS type lattice model considered in
this paper, but can be done in continuous models, we ignore it in the
present study.

Among the new features we find is the existence of a critical end point
(CEP)~\cite{cep}, and the triple point~\cite{tp} in the force versus
temperature phase diagram for sufficiently strong attraction with the
substrate. For weaker attraction, certain phases may not be
thermodynamically stable.  Some of the details of the CEP have been
reported in a shorter paper~\cite{Kapriepl08}. Here we give all the
necessary details and calculations and focus on the triple point. We use
the generating function, the exact transfer matrix and Monte Carlo
simulations to explore the equilibrium behaviour of this DNA-substrate
system under a force on one strand.

The paper is organized as follows: In Sec.~\ref{model}, we define our
model.  Section~\ref{sec:polymer} is devoted to two extreme limits in
which the problem reduces to the unzipping of adsorbed polymer from a
straight and a zig zag hard-wall. These results are used in the
subsequent sections. As a prelude to the substrate interaction problem
in hand, we also need to consider the case of unzipping of a dsDNA
strand. In Sec.~\ref{sec:opposite}, we review the unzipping of a dsDNA
by pulling its strands in opposite directions. The unzipping by pulling
a single strand is studied in Sec.~\ref{sec:single}. In
Sec.~\ref{sec:adsDNA}, the unzipping of an adsorbed DNA by pulling a
single strand is studied.  The existence of a triple point~\cite{tp} and
a critical end point are shown here.  There occurrences are not
dependent on each other. These emerges as the relative attraction of the
substrate is made stronger. In addition to providing some details on CEP
this section focuses on the triple point. Finally, we draw our
conclusions in Sec.~\ref{sec:conclusions}.

\section{Model}\label{model}

\figmodel

We model the DNA by two directed self avoiding walks (DSAWs) on a
$D=1+1$ dimensional square lattice. The walks, starting from the origin,
are directed along the diagonal of the square ($z$ direction). The walks
are not allowed to {\it cross each other} but whenever they meet (i.e.
$x_1 = x_2$) there is a gain in energy $-\epsilon_b (\epsilon_b > 0)$
for every contact. At the diagonal ($x=0$) there is an impenetrable
attractive surface, with an energy $-\epsilon_w (\epsilon_w >0)$, which
favors adsorption of the DNA.  In $1+1$ dimensions, the surface is a
line passing through the diagonal of a square lattice, and only one of
the strands can get adsorbed on it (i.e. $x_1 = 0$), since the two
strands of the DNA cannot cross each other. One end of the DNA is always
kept anchored at the origin. We apply an external force $g$, along the
transverse direction ($x$-direction) on the free end of one of the
strands of the DNA.  The other strand is left free.  Henceforth, the
strand which is left free is called the ``free strand" and the strand,
on which the external force acts is called the ``pulled strand". A
schematic diagram of the model in $1+1$ dimensions is shown in
Fig.~\ref{fig:model}. In $2+1$ dimensions, the surface is a plane
passing through the diagonal of a cubic lattice.  Unlike the $1+1$
dimensional case, both the strands of the DNA can get adsorbed on the
surface and still satisfy the non-crossing constraint on the plane
($y$-direction).  In this paper we concentrate only on $1+1$ dimensional
case.

Let $D_{n}(x_1,x_2)$ be the partition function (temperature dependence
not shown explicitly) of a dsDNA in the fixed distance ensemble where
$n$th monomers of the strands are at positions $x_1$ (free strand) and
$x_2 \ (x_2 \ge x_1)$ (pulled strand) respectively from the wall.
$D_{n}(x_1,x_2)$ satisfies the recursion relation
\begin{eqnarray}\label{recpfn}
	D_{n+1}(x_1, x_2) &=& \sum_{i,j=\pm 1} D_{n}(x_1 + i, x_2 + j)
	\nonumber \\
	& & \times \left[1 + {\mathcal W}\delta_{x_1,0}\right] \left[1
	+ {\mathcal B}\delta_{x_2,x_1}\right],
\end{eqnarray}
where ${\mathcal W} = (e^{\beta \epsilon_w} - 1)$, ${\mathcal B} =
(e^{\beta \epsilon_b} - 1)$ and $\beta = 1/T$ is the inverse temperature
in units of $k_{\rm B}=1$, with the initial condition $D_{0}(x_1, x_2) =
(e^{\beta \epsilon_w} \delta_{x_1,0})(e^{\beta \epsilon_b}
\delta_{x_2,0})$.  The impenetrability of the surface and the
non-crossing of the strands demand $x_2 \ge x_1 \ge 0$.  The canonical
partition function with an external force $g$ at the end of the pulled
strand is then obtained by summing over all the allowed configurations
of the DNA of length $N$ on the lattice.
\begin{equation}\label{recpfnff}
	Z_N(\beta, g) = \sum_{x_2 \ge x_1 \ge 0}  D_N(x_1, x_2) \
	e^{\beta g x_2},
\end{equation}
where $e^{\beta g x_2}$ is the Boltzmann weight due to the force $g$.

From $x_1$ and $x_2$, one can define the relative coordinate $x$, and
the center of mass coordinate $X$ as  
\begin{equation} \label{eq:coord}
	x = x_2 - x_1, \qquad	X = \frac{1}{2} \left[ x_1 + x_2 \right].
\end{equation}

The relative coordinate contains all the necessary information of
the unzipping transition of a dsDNA when its strands are pulled in the
opposite directions~\cite{somen:unzip,maren,scaling,Lubensky}. For the
single strand pulling case we need both the relative and the center of
mass coordinate to track the chains individually. 

\subsection{Quantities of Interest}

The following physical quantities are of interest:

\begin{enumerate}
	\item The average distances of the end monomers of both strands from the
surface, which are defined as
\begin{equation}\label{eq:x12}
	\langle x_{i} \rangle = \sum_{x_2 \ge x_1 \ge 0} \frac{
	x_{i} \ D_N(x_1, x_2) \ e^{\beta g x_2}}{Z_N(\beta, g)},
\end{equation}
where $\langle \cdots \rangle$ denotes the thermal averaging and
subscript $i=1$ and $i=2$ stand for the free and the pulled strand
respectively. The summation is taken over all the allowed configurations. 
The positions of end monomers give us the information needed to
characterize the phase of the system. 

\item The response to the force, i.e. the isothermal extensibility, which
can be expressed, for both strands, in terms of fluctuations of the
position of end monomers
\begin{equation}\label{eq:ext}
	\chi_{i} = \left. \frac{\partial \langle x_{i} \rangle
	}{\partial g} \right |_{T} =  \frac{1}{k_B T} \left [ \langle
	x_{i}^{2} \rangle - {\langle x_{i} \rangle}^2 \right ], \quad (i=1~
	\text{or}~2).
\end{equation}
\end{enumerate}
Equations~(\ref{eq:x12}) and (\ref{eq:ext}) are useful to obtain the
above physical quantities numerically.  

\section{Unzipping an Adsorbed Polymer}\label{sec:polymer}

Let us first study the two extreme limits: (i) $\epsilon_{\rm b}
\rightarrow \infty$ and $\epsilon_{\rm w}$ is finite (and we take
$\epsilon_{\rm w} = 1$), and (ii) $\epsilon_{\rm w} \rightarrow \infty$
and $\epsilon_{\rm b}$ is finite ($\epsilon_{\rm b} = 1$). For the
former case, the two strands of the DNA always stay together for the
entire range of temperature in which we are interested. In such a
situation, an external force $g$ on the pulled strand also pulls the free
strand. Therefore, the DNA can be equivalently represented by a single
polymer in the center of mass frame. The problem then reduces to the
unzipping of an adsorbed polymer on an straight impenetrable surface
(hard-wall). However, for case (ii), the free strand remains adsorbed on
the surface and itself acts as a zig zag hard-wall for the pulled
strand. In this section, we obtain the phase diagram for both the cases. 

\subsection{Straight hard-wall} \label{sec:straight}

In the fixed distance ensemble, the partition function $w_n(x)$ of a
dsDNA in the center of mass frame satisfies the recursion relation
\begin{equation}\label{recst}
	w_{n+1}(X) = \left[ w_{n}(X+1) + w_{n}(X -1) \right] \left[ 1 +
	{\mathcal W} \delta_{X,0} \right].
\end{equation}
Note that while writing Eq.~(\ref{recst}) from Eq.~(\ref{recpfn}) we
have absorbed the Boltzmann factor $e^{\beta \epsilon_{\rm b}}$ in
$w_n(X)$. 
The partition function in the presence of an applied external
force $g$ is then given by
\begin{equation}
	Z_N(\beta, g) = \sum_{X \ge 0} w_N(X) \ e^{\beta g X}.
\end{equation}
This problem has been studied in recent years
because of its similarity with the DNA
unzipping~\cite{kapri:pre05,orlandini}. It is known that the polymer
unzips from the surface if the force exceeds a critical value. Below
this critical force, the polymer remains adsorbed on the surface while
above it the polymer is in the desorbed phase. The phase boundary
separating the two phases is given by
\begin{equation} \label{eq:gcpoly}
	g_{\rm c}^{(s)}(T) = \frac{T}{2} \ln \left[ e^{\beta \epsilon_{\rm
	w}} - 1 \right].
\end{equation}
and is shown in Fig.~\ref{phzig} by a solid line. The triangles on the
line represent the phase boundary obtained numerically by using the
exact transfer matrix. The method is introduced below for the zig zag
surface. Here we have taken $\epsilon_{\rm w}=1$. 
The critical force decreases monotonically with the increase of
temperature and becomes zero at 
\begin{equation}
	T_{\rm c}^{(s)} = \epsilon_{\rm w}/\ln{2},
\end{equation}
where the polymer desorbs from the surface because of the thermal
fluctuations.

\figphzig

\subsection{Zig zag surface} \label{sec:zig}

The recursion relation for the partition function $b_n(x_2)$ of the
pulled strand, in the fixed distance ensemble, can be written as
\begin{eqnarray} \label{reczz}
	b_{n+1}(x_2) &=& \left[ b_n(x_2 + 1) + b_n(x_2 - 1) \right]
	\nonumber \\
	& & \times \left[ 1 + {\mathcal B} \delta_{x_2,0} \right] \left[ 1 +
	{\mathcal B} \delta_{x_2,1} \right].
\end{eqnarray}
Note that the above recursion relation differs from Eq.~(\ref{recst}) in
the extra factor due to the binding of the odd monomers to the wall. In
Eq.~(\ref{recst}), only the even monomers bind to the surface. This
modification is enough to make a difference in the phase
diagram. This extra energy affects the intermediate temperature
behaviour of the polymer as it becomes energetically costly to create
bubbles on the polymer adsorbed on a zig zag hard-wall. This can be
understood as follows: The ground state for the polymer adsorbed on a
straight hard-wall contains inherent bubbles of length $\ell=2$ because the
geometry of the problem allows only $N/2$ monomers on the wall. In
contrast, for the zig zag hard-wall there are no such inherent bubbles
as all the $N$ monomers are adsorbed on the wall.  To create a bubble of
length $\ell=2$, one of the monomers has to come out from the wall which
costs energy. Furthermore, getting larger bubbles are easier for the
polymer adsorbed on the straight hard-wall than the zig zag case.  For
example, flipping of one monomer from the wall can create a bubble of
length $\ell=4$ for the normal hard-wall, while three consecutive monomers
have to come out from the wall for the zig zag case to create a bubble
of the same size. 

Under the influence of a fixed pulling force $g$ at the free end, the
partition function is obtained by
\begin{equation}\label{reczzff}
	{\mathcal Z}_N(\beta, g) = \sum_{x_2} b_N(x_2) e^{\beta g x_2}.
\end{equation}

The partition function of the chain length $N$ is obtained numerically,
at a given temperature, by iterating Eq.~(\ref{reczz}) and, then, for
various force $g$ by using Eq.~(\ref{reczzff}). This is known as the
exact transfer matrix technique. The distance of the end monomer of the
pulled strand from the surface, which is the quantity of interest, is
obtained by using
\begin{equation}
	\langle x_2 \rangle = \frac{1}{ {\mathcal Z}_N(\beta, g)} \sum_{x_2}
	x_2 b_N(x_2) e^{\beta g x_2}.
\end{equation}

\figxgzig

In Fig.~\ref{xgzig}(a), we have shown the data collapse of $g$ vs
$\langle x_2 \rangle$ isotherms at $T=1.0$ for the chain of lengths
$N=1000$, $2000$ and $3000$. We have used the scaling form
\begin{equation}
	\langle x_2 \rangle = N^{d} {\mathcal Y} \left( (g - g_{\rm
	c}^{(z)}) N^{\phi} \right),
\end{equation}
where $d$ and $\phi$ are the critical exponents and $g_{\rm c}^{(z)}$ is
the critical force. By using the Bhattacharjee-Seno
procedure~\cite{bh:seno}, we obtained $g_{\rm c}^{(z)} = 1.059 \pm
0.001$, $d = 0.99 \pm 0.01$ and $\phi = 1.0 \pm 0.01$. These exponents
are same as the exponents obtained for the unzipping from the straight
hard-wall~\cite{kapri:pre05} and the DNA unzipping
problem~\cite{maren,scaling}. We use same procedure at various
temperatures to obtain the phase diagram.

The phase diagram of unzipping from the zig zag surface is shown in
Fig.~\ref{phzig} by points for $\epsilon_b = 1$. If we compare this
with the phase boundary for the straight surface, we see that at very
low temperatures ($T=0$), the force required to unzip the polymer from the
zig zag surface is twice that from the normal surface since one has to
overcome an additional binding at $x_2=1$. This implies the low
temperature behavior of the polymer is same for both the walls i.e. $g
\propto \epsilon$ ( $\epsilon = \epsilon_w$ for the straight surface
whereas $\epsilon = 2 \epsilon_b$ for the zig zag case). The polymer
starts behaving differently as temperature is increased.  As already
mentioned, it is easy to create a bubble on the polymer adsorbed on the
straight wall, whereas more energy is needed to create a bubble of the
same size on the polymer adsorbed on the zig zag surface.  For the
straight surface, the critical force needed to unzip the polymer from
the wall decreases monotonically with the temperature.  In contrast, for
the zig zag surface, the critical force increases at intermediate
temperatures, reaches to a maximum value and then decreases to zero,
thus showing reentrance as shown in the inset of Fig.~\ref{phzig}. To
obtain the desorption temperature for the zig zag case, we resort again
to the finite size scaling (but with temperature as a variable) of the
form
\begin{equation}\label{fsstm}
	\langle x \rangle = N^{d_t} {\mathcal Y}\left( (T - T_{\rm c}^{(z)})
	N^{\phi_t} \right),
\end{equation}
with $T_{\rm c}^{(z)}$ as the critical desorption temperature and $d_t$
and $\phi_t$ are the critical exponents. The data collapse, for chain
lengths $N=2000$, $3000$, $4000$ and $5000$, obtained for $d_t = 0.53
\pm 0.03$, $\phi_t = 0.4 \pm 0.03$ and $T_{\rm c}^{(z)} = 4.81 \pm 0.16$
is shown in Fig.~\ref{xgzig}(b), and the desorption temperature $T_{\rm
c}^{(z)}$ is shown in Fig.~\ref{phzig} by a solid circle.

\section{Unzipping DNA by pulling strands in opposite directions}
\label{sec:opposite}

\figphdna

Before considering the unzipping by pulling a single strand, let us
first concentrate on the unzipping of a dsDNA by pulling its strands in
opposite direction by an external force $g$. This problem has
received a lot of attention in recent years due to its resemblance with
the way the DNA unzipping experiments are done~\cite{Roulet,Danilowicz}. 

In the fixed distance ensemble, the partition function of the dsDNA can
be obtained by assigning $\epsilon_{\rm w} = 0$ in Eq.~(\ref{recpfn}).
The recursion relation, in the relative coordinate, reads
as~\cite{scaling}
\begin{equation}
	d_{n+1}(x) = \left[ d_n (x+1) + 2 d_n (x) + d_n (x-1) \right]
	\left[ 1 + {\mathcal B} \delta_{x,0} \right],
\end{equation}
with the initial condition $d_{0}(x) = e^{ \beta \epsilon_{\rm b}}
\delta_{x, 0}$.

In the fixed force ensemble, the partition function for the DNA of
length $N$, is then obtained by
\begin{equation}
	Z_N(\beta, g) = \sum_{x \ge 0} d_N(x) \ e^{\beta g x}.
\end{equation}

The phases of the DNA and the transition come from the singularities of
the of the generating function
\begin{equation}
	{\mathcal G}(z, \beta, g) = \sum_{N = 0}^{\infty} z^N Z_N(\beta, g).
\end{equation}
The singularities are~\cite{maren,scaling}
\begin{subequations}
	\begin{equation}\label{eq:z1}
		z_1 = \frac{1}{4},
	\end{equation}
	\begin{equation} \label{eq:z2}
		z_2(\beta, \epsilon_{\rm b}) = \sqrt{ 1 - e^{-\beta
		\epsilon_{\rm b}} } - 1 + e^{-\beta \epsilon_{\rm b}},
	\end{equation}
	and
	\begin{equation}\label{eq:z3}
		z_3(\beta, g) = \frac{1}{2 + 2 \cosh \beta g}.
	\end{equation}
\end{subequations}
The phase of the DNA is given by the singularity which is closest to the
origin and the phase transition takes place when the two singularities
cross each other. For low force, $z_2(\beta, \epsilon_{\rm b})$ is
closest to the origin and the DNA is in the zipped phase
(double-stranded), while for high force, $z_3(\beta, g)$ becomes closest
and the DNA is in the unzipped phase (two single strands). The
force-temperature phase boundary between the two phases is obtained by
equating the two singularities, which gives~\cite{scaling,KapriSMB}
\begin{eqnarray}\label{eq:gcffe}
	g_{\rm c}(T, \epsilon_{\rm b}) &=& \frac{T}{2} \cosh^{-1} \left [
	\frac{1}{2 z_2(\beta, \epsilon_{\rm b})} -1 \right ] \nonumber\\
	&=& - \frac{T}{2} \ln \lambda \left( z_2(\beta, \epsilon_{\rm b}) \right),
\end{eqnarray}
where $\lambda(z) = (1-2z -\sqrt{1-4z})/(2z)$. The thermal denaturation
(melting), coming from $z_1 = z_2(\beta, \epsilon_{\rm b})$, is at 
\begin{equation}
	T_{\rm m} = \epsilon_{\rm b} / \ln (4/3).
\end{equation}
The phase boundary separating the zipped phase (Zp) from the unzipped
phase (Uz) is shown by the dashed line in Fig.~\ref{fig:phdna}.


\section{Unzipping DNA by pulling a single strand} \label{sec:single}

In previous sections, we have seen that the finite size scaling of the
force-distance isotherms (or the extensibility), obtained by using the
exact transfer matrix for various chain lengths, can be used to obtain
the phase diagrams numerically. We use Eqs.~(\ref{recpfn}) and
(\ref{recpfnff}) with $\epsilon_{\rm w} = 0$ to obtain the partition
function of a dsDNA of length $N$. The average distances of the end
monomers of both the strands from the surface is obtained by using
Eq.~(\ref{eq:x12}).

Alternatively, using the fact that the force is only on one of the
strands and the surface plays no role in the phase boundary, the
partition function can be obtained, in the relative coordinates, by the
following recursion relation (in a mixed fixed-distance-force ensemble)
\begin{widetext}
\begin{equation}
	\label{eq:recmixed}
	d_{n+1}(x) = \left[ d_{n}(x+1) \ e^{\beta g} + d_{n}(x) \ e^{\beta g} +
	d_{n}(x) \ e^{-\beta g} + d_{n}(x-1) \ e^{-\beta g} \right] \left[ 1 +
	{\mathcal B} \delta_{x,0} \right],
\end{equation}
\end{widetext}
with the initial condition $d_{0}(x) = e^{\beta \epsilon_{\mathrm b}}
\delta_{x,0}$. The above recursion relation can be analyzed analytically
and the boundary separating the zipped and the unzipped phases can be
obtained exactly. The details of the calculation are given in Appendix
A.

\subsection{Isotherms and Extensibility}

\figxgzero

The force-distance isotherms at $T = 0.5$ and $1.5$ are shown in
Fig.~\ref{fig:xg0}(a). Due to the entropic repulsion, the dsDNA stays at a
distance of $\sqrt{N}$ from the surface, even for $g= 0$, to maximize its
entropy (see Fig.~\ref{fig:xg0}(b)).  When the pulling force, $g$, is small,
the binding energy, $\epsilon_{\rm b}$, wins over the entropy, which the free
strand can gain if it separates itself from the pulled strand, and the dsDNA as
a whole gets stretched in the direction of the force. The average distances of
end points of the strands from the surface, $\langle x_{1,2} \rangle/N$, remain
the same (for both the strands) and increase linearly with $g$. The slope,
however, depends on the temperature; it is larger at low temperatures and
smaller at high temperatures.  As $g$ is increased further, the dsDNA gets
completely stretched at $T=0.5$. But for $T=1.5$, before the dsDNA can get
fully stretched, a critical force $g_{\rm u}$ is reached, and the free strand
of the DNA gets unzipped from the pulled strand to increase its entropy. We
call this as ``transition Uz''. This transition can be studied in the relative
coordinate, $x$, defined in Eq.~(\ref{eq:coord}).  Below $g_{\rm u}$, the
strands stay together ($\langle x \rangle / N \rightarrow 0$ as $N \rightarrow
\infty$), and above it, they  are maximally separated ($\langle x \rangle \sim
N$).


The isothermal extensibility, obtained by using Eq.~(\ref{eq:ext}), is
plotted in  Fig.~\ref{fig:ext0}(a) as a function of $g$ for various
chain lengths at $T=1.5$. The critical force $g_{\rm u}$ can be located
by using the finite size scaling of the form 
\begin{equation}
   \label{eq:fss}
	\chi = N^{d} {\mathcal G}\left ( (g - g_{\rm u}) \ N^{\phi} \right ),
\end{equation}
with $d$ and $\phi$ as critical exponents. By using the Bhattacharjee-Seno
procedure, we obtained $d=2.02 \pm 0.01$, $\phi = 1.01 \pm 0.01$ with $g_{\rm
u} = 2.692 \pm 0.001$. These exponents are same as the exponents for 
the DNA unzipping by pulling strands in the opposite
directions~\cite{maren,scaling,KapriSMB:JPCM}. This indicates that, similar to the
later case, the unzipping by pulling a single strand is also a first
order phase transition with
\begin{equation}
	\chi / N ~\sim~ \mid g - g_{\rm u} \mid^{-1}.
\end{equation}
The data collapse is shown in Fig.~\ref{fig:ext0}(b). The finite size
scaling of extensibility, as described above, can be used at various
temperatures to obtain the phase diagram of unzipping.

\figextzero

\subsection{Phase Diagram}

The boundary separating the zipped (Zp) and the unzipped (Uz) phases
is given by (see Appendix A for details)
\begin{equation}
	\label{eq:gcsingpull}
	g(T) = \frac{k_{\mathrm B}T}{2} \ln \left( \frac{ 2 e^{-\beta
	\epsilon_{\mathrm b}} - 2} { 1 - 2 e^{-\beta \epsilon_{\mathrm b}}}
	\right).
\end{equation}
The same phase boundary has also been obtained by Marenduzzo {\it et
al.}~\cite{maren2008} in a different context of melting of a stretched
DNA.  The phase boundary is shown in Fig.~\ref{fig:phdna} by a solid
line which matches excellently with the results (triangles) obtained
numerically from the exact transfer matrix. The plot shows that in
contrast to pulling both the strands in opposite directions, where,
below the melting temperature, the DNA can be unzipped at all
temperatures including $T=0$, the DNA can only be unzipped above a
certain temperature $T_{\rm u}$ ($T_{\rm u} = \epsilon_{\rm b}/\ln 2$)
for the single strand pulling. Below $T_{\rm u}$, the DNA remains in the
zipped phase for any value of force $g$. The melting temperature of the
dsDNA remains the same, i.e. $T_{\rm m}=\epsilon_{\rm b} / \ln(4/3)$,
because in our model the force acts in the transverse direction and does
not overstretch the DNA. Recently it was found that a longitudinal
stretching force on one of the strand destabilize the DNA which results
in a reduced melting
temperature~\cite{Rouzina01,Hanke08,Rudnick08,Rahi08}.

The origin of the unzipping transition by pulling a single strand is
different from the unzipping transition by pulling both the strands. For
the latter case, the transition sets in due to the competition between
the base pairing energy $\epsilon_{\rm b}$, which binds the
complementary bases (or monomers) of two strands, and the orientation of
the individual links connecting the monomers.  In contrast, the
interplay between the binding energy $\epsilon_{\rm b}$, and the
entropy, which the free strand can gain if it is in the unzipped phase,
is responsible for the single chain case. When $T < T_{\rm u}$, the
binding energy wins over the entropy and the DNA remains in the zipped
phase for any value of $g$. For large $g$, the DNA takes a fully
stretched configuration and bubbles are not possible. This is analogous
to the ${\mathsf Y}$ model studied in Ref.~\cite{scaling}. In this
model, the thermal melting of dsDNA takes place at $T_{u} =
\epsilon_{\rm b} / \ln 2$ and the transition is of first
order~\cite{scaling}. For $T < T_{\rm u}$, the DNA always remains in the
zipped phase. For $T > T_{\rm u}$, the free strand has all the
favourable conditions to increase its entropy. Therefore, as $T$ is
increased, the free strand can get separated from the pulled strand well
below the fully stretched configuration and the critical force $g_{\rm
u}$ falls rapidly with increasing the temperature, becoming zero at
$T_{\rm m}$. 

There are other factors that also contribute in the unzipping of dsDNA
by pulling a single strand but neglected in our lattice model. An
important one is the fact that the length per base pair of the dsDNA and
the ssDNA are different (0.34nm and around 0.5mn respectively).  If the
applied force is low, the dsDNA has larger extension than the ssDNA and
the dsDNA remains stable.  However, for sufficiently strong force, the
extension of ssDNA becomes more than the dsDNA and the force destabilize
the dsDNA and favors its unzipping~\cite{Rouzina01}. It is difficult to
incorporate this in a lattice model like ours, but we believe that it
can be done in a continuous description of the model. It would be
interesting to study the combined effect of both mechanisms in DNA
unzipping by pulling a single strand as this could bring down $T_{\rm
u}$.  

\section{Unzipping an adsorbed DNA} \label{sec:adsDNA}

Let us now consider the complete model of an attractive surface (i.e.,
$\epsilon_{\rm w} > 0$) near a dsDNA. Here the free strand of the DNA
experiences two different energies of opposite tendencies even at $T=0$.
The energy $\epsilon_{\rm w}$ tries to keep the free strand adsorbed on
the surface while the binding between the strands, $\epsilon_{\rm b}$,
tries to keep it with the pulled strand. For $T>0$, the entropy also
plays a role and the competition among them makes the phase diagram very
rich. We establish the possibility of four distinct phases (I) \phI:
zipped DNA adsorbed on the surface, (II) \phII: zipped DNA desorbed from
the surface, (III) \phIII: unzipped DNA with the free strand adsorbed on
the surface, and (IV) \phIV: unzipped DNA with both the strands desorbed
from the surface.

\subsection{Isotherms and Extensibility}

\figxg

In Fig.~\ref{fig:xg}, we have shown the force-distance isotherms for
$\epsilon_{\rm w} = 1.0$, $1.8$ and $2.0$ at two different temperatures
$T = 0.5$ and $1.5$ for the chain of length $N=1000$. The values of
$\epsilon_{\rm w}$ and $T$ are selected to display the typical
characteristics of the phase diagrams. The phase diagrams are obtained
by repeated use of finite size scaling of the response function.

For $\epsilon_{\rm w} = 1.0$, the isotherms are shown in
Fig.~\ref{fig:xg}(a).  In the absence of a force $g$, the ground state
is an adsorbed DNA on the surface. As $g$ is increased, we see that
there is a critical force, $g_{\rm s}$, at which the DNA gets unzipped
from the surface but remains double stranded. We call this as
``transition Sz''.  This involves the center of mass coordinate, $X$,
given by Eq.~(\ref{eq:coord}).  Below $g_{\rm s}$, $\langle X \rangle /
N = 0$ as $N \rightarrow \infty$, but it takes a finite (non-zero) value
above $g_{\rm s}$. When $g$ is increased further, the dsDNA stretches
more and more in the direction of the force, and takes a fully stretched
configuration for $T=0.5$.  However, for $T=1.5$, the free strand gets
separated from the pulled strand (transition Uz) at $g_{\rm u}$ and
stays at a distance of $\sqrt{N}$ from the surface (see
Fig.~\ref{fig:xg}(a)).

The isotherms for $\epsilon_{\rm w} = 1.8$ is shown in
Fig.~\ref{fig:xg}(b). At $T=0.5$, the isotherm is similar to that of
$\epsilon_{\rm w} = 1$, though with a higher critical force $g_{\rm s}$.
It is different for $T=1.5$, where, on increasing the force $g$, the
transition Sz takes place at $g_{\rm s}$. On increasing $g$ further, we
have an another transition Uz at $g_{\rm u}$, where the free strand
separates itself from the pulled end and gets adsorbed on the surface.
We have obtained the isotherm at $T=1.5$ by two different methods. The
smaller symbols with broken lines between them are from the exact
transfer matrix whereas, the bigger symbols (squares and circles
representing the two strands of the DNA) are obtained by performing
Monte Carlo simulations, at their respective $g$ values. We also collect
histograms, $h_k(E, x_1, x_2)$ ($E$ is the total binding energy, $x_1$
and $x_2$ are respectively the distances, from the surface, for the free
and the pulled strand, and the subscript $k$, stands for the simulation
performed at force $g_k$) at each simulation.  These histograms are then
combined to estimate $\langle x_{1,2} \rangle$ for a range of forces by
using the multiple histogram technique~\cite{ferren89}. The estimates,
so obtained, are shown by the upper and the lower triangles for the free
and the pulled strand respectively in Fig.~\ref{fig:xg}(b). The isotherm
obtained from both the methods agree excellently. The details of the
Monte Carlo simulation is discussed in Appendix B. Our Monte Carlo
simulations in $2+1$ dimensions also shows similar
results~\cite{Kapriepl08}.

The isotherms for $\epsilon_{\rm w} = 2$ are shown in
Fig.~\ref{fig:xg}(c).  The plot shows that the only possible transition
is the unzipping of the dsDNA to two single strands (i.e. the transition
Uz) at $g_{\rm u}$, because the free strand minimizes its energy by
staying adsorbed on the surface. 

The critical force $g_{\rm s}$ at which the transition Sz takes place is
obtained by the finite size scaling of isothermal extensibility as given
by Eq.(\ref{eq:fss}). A good data collapse is obtained for the exponents
$d_{\rm s} = 1.95 \pm 0.05$ and $\phi_{\rm s} = 1.0 \pm 0.01$ showing
that, as for transition Uz, the transition Sz is also first order. We
use the finite size scaling of extensibility at various temperatures to
obtain the phase diagrams.

\subsection{Phase Diagram}

The discussions in the previous subsection reveal that there are four possible
phases for the problem at hand. These are: 

\begin{enumerate}

	\item {\it Phase I} --- The zipped DNA (i.e., dsDNA) adsorbed on the
		surface. This phase is characterized by $\langle x_{1,2}
		\rangle/N \rightarrow 0$ as $N \rightarrow \infty$. {\it This
		phase is called}~\phI.

	\item {\it Phase II} --- The dsDNA desorbed from the surface. In
		this phase we have $\langle x_{1,2} \rangle/N \rightarrow A \ (
		0 < A 	\le 1)$ as $N \rightarrow \infty$. {\it This phase is
		called}~\phII.

	\item {\it Phase III} --- In this phase, the DNA is unzipped (i.e.,
		the strands stay away from each other). The pulled strand is
		stretched in the direction of the force and the free strand is
		adsorbed on the surface. This phase is characterized by $\langle
		x_1\rangle/N \rightarrow 0$ but $\langle x_2 \rangle /N
		\rightarrow 1$ as $N \rightarrow \infty$. {\it This phase is
		called}~\phIII.

	\item {\it Phase IV} --- In this phase both the strands stay away
		from the surface as well as from each other.  The pulled strand
		follows the pulling force as in {\it phase III}, and the free
		strand stays away from the surface to maximize its entropy. This
		phase is characterized by $\langle x_2 \rangle /N \rightarrow 1$
		and $\langle x_1 \rangle/\sqrt{N} \rightarrow 1$ as $N
		\rightarrow \infty$. {\it This phase is called}~\phIV.

\end{enumerate}

Depending on the relative strength of the binding energy, $\epsilon_{\rm
w}$, and the pairing energy $\epsilon_{\rm b}$, we can either have all
the four phases listed above, or a few of them, in the phase diagram. We have
four parameters: $g$, $T$, $\epsilon_{\rm w}$ and $\epsilon_{\rm b}$.
Out of them we can construct three dimensionless quantities $g
\rightarrow g/\epsilon_{\rm b}$, $T \rightarrow k_{\rm B}T/\epsilon_{\rm
b}$ and $ \epsilon_{\rm w} \rightarrow \epsilon_{\rm w}/\epsilon_{\rm
b}$. Therefore, without loss of generality, we chose $\epsilon_{\rm b} = 1$.
The phase of the DNA, for the given set of parameters, can be read from
a 3-dimensional $g$-$T$-$\epsilon_{\rm w}$ surface. Since it is
difficult to show a 3-dimensional plot, we show the cross-sections
($g$-$T$ plane) of the above surface for various $\epsilon_{\rm w}$.
These are shown in Fig.~\ref{fig:phdia}.

Before discussing the phase diagrams in detail, let us do a zero
temperature ($T=0$) analysis of the problem, keeping $\epsilon_{\mathrm
b}$ constant. The energies of the three phases, namely \phI, \phII, and
\phIII ~are respectively given by
\begin{subequations}
	\begin{equation}
		E_{{\mathrm Z_a}} = -N(\epsilon_{\mathrm w}/2 + 1),
	\end{equation}
	\begin{equation}
		E_{\mathrm{Z_d}} = -N(g + 1),
	\end{equation}
	and
	\begin{equation}
		E_{{\mathrm U}_{\mathrm ad}} = -N( \epsilon_{\mathrm w}/2 + g).
	\end{equation}
\end{subequations}
For $1 < \epsilon_{\mathrm w} < 2$, the phase \phIII ~is always unstable
(i.e. it has higher energy). The transition from phase \phI ~to phase
\phII ~occurs at $g = \epsilon_{\mathrm w} / 2$. But for
$\epsilon_{\mathrm w} > 2$, \phII ~is not possible. At
$\epsilon_{\mathrm w} = 2$, there is a degeneracy for \phII ~and \phIII,
which occurs at $g = \epsilon_{\mathrm w}/2$. Therefore at this point
all the three phases coexist and it is a triple point.

\subsubsection{ $ 0 < \epsilon_{\rm w} \le 1$ and $1 <
\epsilon_{\rm w} < 2$}

When $0 < \epsilon_{\rm w} \le 1$, the phase diagram contains three
phases, namely \phI, \phII ~and \phIV. But, for $1 < \epsilon_{\rm w} <
2$, a new phase, \phIII, also appears in the phase diagram.  The phase
boundary separating \phI ~from \phII ~which decreasing monotonically for
$\epsilon_{\rm w} < 1$ now shows reentrance at intermediate
temperatures. Apart from this feature, there is a region in the phase
diagram which involves three phases, namely \phII, \phIII ~and \phIII.
The transition from phase \phII ~to \phIII ~and \phIV ~are of first
order, whereas, the transition from phase \phIII ~to phase \phIV ~is a
second order. The second order phase boundary terminates on the first
order line at a critical end point. These are discussed in
Ref.~\cite{Kapriepl08}.

\figphdia

\subsubsection{ $\epsilon_{\rm w} = 2$: triple point and critical end
point }

The phase diagram for $\epsilon_{\rm w} = 2$ is shown in
Fig.~\ref{fig:phdia}. It contains all the four phases, namely \phI,
\phII, \phIII, and \phIV, same as $1 < \epsilon_{\rm w} < 2$ case.  The
phase boundary separating phase \phI ~from phase \phII ~is shown by
circles.  For comparison, we have also shown, by the dashed line, the
phase boundary (see Eq.~(\ref{eq:gcpoly})) for the unzipping of an
adsorbed polymer from the straight hard-wall. At low $T$, the two
strands of the DNA, which are bound to each other, behave like a single
chain, and therefore, the low-$T$ phase boundary for both the problems
match with $g(T=0) = \epsilon_{\rm w}/2$.  As $T$ is increased, bubbles
form on the DNA.  These bubbles create an entropic hindrance on the free
strand to stay away from the surface. As a result desorption of dsDNA
from the surface occur at a temperature higher than the temperature
needed to desorb a single polymer i.e. $T_{\rm w} > \epsilon_{\rm w} /
\ln 2$. The boundary separating phase \phII ~with phase \phIV, which is
shown by a solid line, is again given by Eq.~(\ref{eq:gcsingpull}),
i.e., it is same as the phase boundary between the zipped and the
unzipped phases for $\epsilon_{\mathrm w} = 0$ case. Therefore, in this
region, the surface plays no role in determining the phase boundary
between the zipped and the unzipped phases. The transition from phase
\phII ~to \phIII, which is well approximated by 
\begin{equation}
	\label{eq:gcsurf}
	g(T) = \frac{k_{\mathrm B} T}{2} \ \ln \left( \frac{ 2 \sqrt{
	e^{-\beta \epsilon_{\mathrm w}} - e^{-2 \beta \epsilon_{\mathrm w}}
	} - e^{-\beta \epsilon_{\mathrm b}} } { e^{-\beta \epsilon_{\mathrm
	b}} - \sqrt{ e^{-\beta \epsilon_{\mathrm w}} - e^{-2 \beta
	\epsilon_{\mathrm w}}}} \right),
\end{equation}
for the $\epsilon < 2$ case (see Appendix A and Ref.~\cite{Kapriepl08})
is shown by the dotted lines. This curve deviates markedly with the data
obtained by the exact transfer matrix calculations at intermediate
temperatures, but still matches in both the high and the low
temperatures. This deviation is because we can no longer neglect the
effect of surface on the pulled strand mediated via the free strand and
the approximation on which the Eq.~(\ref{eq:gcsurf}) is based breaks
down. At low temperature side, it meets with the dashed line at $T=0$,
showing that the three phases coexist there as expected from zero
temperature analysis discussed above. This is the triple point ($T_{\rm
p} = 0$) which is shown by ${\mathrm C}_{\rm p}$ (filled diamond) in the
phase diagram. The numerics, however, cannot resolve the two phase
boundaries at low temperatures. On the high temperature side, the dotted
curve meets with the solid line with the same slope but with different
curvature at the CEP which is shown by ${\mathrm C}_{\rm e}$ (filled
square). At this point the boundary separating phases \phIII ~and \phIV
~gets terminated.

\subsubsection{ $\epsilon_{\rm w} > 2$}

When $\epsilon_{\rm w}$ is slightly greater than $2$, the triple point
is at a finite temperature ($T_{\rm p} > 0$). However, it is difficult
to get the exact location of the triple point because of the difficulty
in resolving the phase boundaries at low temperatures.  On increasing
$\epsilon_{\rm w}$ further, the  region representing phase \phII
~shrinks rapidly and both the triple point and the CEP shift towards
higher temperature and disappear independently from the phase diagram.
When $\epsilon_{\rm w} = \infty$, the free strand, which remains
adsorbed on the surface at all temperatures, acts like a zig zag
hard-wall studied in Sec.~\ref{sec:polymer}. In this case, only the
phase \phI ~and the phase \phIII ~survive in the phase diagram as shown
in Fig.~\ref{phzig}. This picture is true only in $D=1+1$ dimensions. In
$D=2+1$ dimensions the winding of chains will produce dynamic hindrance
to unzipping unless torque releasing mechanism is provided at the
anchored point. Same will be true for a real DNA, where its helical
structure will prevent its unzipping though topoisomerases may help in
releasing the extra strain. There is also the possibility of dynamic
hindrance leading to nonequilibrium long lived states. These are beyond
the scope of this work.

\section{Conclusions} \label{sec:conclusions}

In this paper, we have studied the unzipping of a dsDNA by pulling a
single strand. We find that a dsDNA can be unzipped, even by pulling a
single strand, if the pulling force exceeds the critical value. The
origin of this transition is different from the unzipping of DNA by
pulling its strands in opposite directions.  In contrast to the latter
case, where the unzipping is possible at all temperatures below the
denaturation temperature of the DNA, the unzipping only starts above
some minimum temperature ($T_{\rm u} = \epsilon_{\rm b}/ \ln 2$) for the
single strand pulling but takes place up to the denaturation
temperature. For both the cases, the transition is first order.   

On introducing the binding energy, $\epsilon_{\rm w}$, at the surface,
the phase diagram becomes very rich, with a total of four possible
phases. Depending upon the relative strength of $\epsilon_{\rm w}$ and
$\epsilon_{\rm b}$, we can either have all the four phases, or a few of
them, in the phase diagram. We find that for a wide range of
$\epsilon_{\rm w}$, there is a critical end point in the phase diagram,
where a line of second order (critical line) terminates on a first order
phase boundary. Furthermore, for a narrow range of $\epsilon_{\rm w}$,
we have a triple point in the phase diagram. As $\epsilon_{\rm w}
\rightarrow \infty$, the problem reduces to the unzipping of a polymer,
which is adsorbed on a zig zag hard-wall. It seems that the unzipping of
an adsorbed dsDNA by pulling a single strand can be a potential
candidate to explore the critical end point in single molecule
experiments. 

\renewcommand{\theequation}{A-\arabic{equation}}
\setcounter{equation}{0}  

\section*{APPENDIX A: EXACT PHASE BOUNDARIES } 

In this appendix we give details of the analytical calculation for
obtaining the phase boundaries by using the recursion relation given 
by Eq.~(\ref{eq:recmixed}). We define the generating function of the
partition function $d_{n}(x)$ by 
\begin{equation}
	\label{eq:A2}
	G(z,x) = \sum_{n = 0}^{\infty} z^n d_{n}(x), 
\end{equation}
which can be taken of the form (ansatz)
\begin{equation}
	\label{eq:A3}
	G(z,x) = \lambda^{x}(z) A(z), 
\end{equation}
where $\lambda(z)$ and $A(z)$ need to be determined. Using this ansatz
in Eq.~(\ref{eq:recmixed}) ($z$ dependence of $\lambda$ and $A$
suppressed) we get
\begin{subequations}
	\begin{equation}
		\label{eq:A4:a}
		\frac{A}{z} = \left[ \left \{ e^{\beta g} + e^{-\beta g} \left(
		1 + \lambda \right) \right \} A + \frac{1}{z} \right] e^{\beta
		\epsilon_{\mathrm b}} \quad \text{for} \ x=0,
	\end{equation}
	and
	\begin{equation}
		\label{eq:A4:b}
		\frac{\lambda^{x}}{z} = \lambda^{x-1} \left(1 + \lambda \right)
		\left[ e^{\beta g}  + \lambda  e^{-\beta g} \right] \quad
		\text{for} \ x > 0,
	\end{equation}
\end{subequations}
from which one obtains
\begin{subequations}
	\begin{eqnarray}
		\label{eq:A5:a}
		\lambda &=&  \frac{1}{2z} \Big[ e^{\beta g} - \left( e^{2\beta
		g} + 1 \right)z \nonumber \\
		&+&	\sqrt{ \left\{ \left( e^{2\beta g} + 1 \right) z - e^{\beta
		g} \right\}^2 - 4z^2 e^{2\beta g} } \Big],
	\end{eqnarray}
	and
	\begin{equation}
		\label{eq:A5:b}
		A = \frac{ e^{\beta \epsilon_{\mathrm b}}}{ 1 - z \left[
		e^{\beta g} + e^{-\beta g} \left( 1 + \lambda \right) \right]
		e^{\beta \epsilon_{\mathrm b}}}.
	\end{equation}
\end{subequations}
Now from Eqs.~({\ref{eq:A2}}) and ({\ref{eq:A3}}), we have
\begin{equation}
	\label{eq:A6}
	G(z) = \sum_{x} G(z,x) = \sum_{x} \lambda^{x} A = \frac{A}{1 -
	\lambda},
\end{equation}
which has a singularity at $\lambda = 1$. Substituting $\lambda = 1$ in
Eqs.~(\ref{eq:A5:a}) and (\ref{eq:A5:b}), we get singularities of the
unzipped and the zipped phases respectively as
\begin{subequations}
\begin{eqnarray}
	\label{eq:A7:a}
		z_4 &=& \frac{1}{4 \cosh{\beta g}} \nonumber \\
		    &=& \left( \frac{1}{2} \right) \left( \frac{1}{2 \cosh \beta
			g} \right),
\end{eqnarray}
and
\begin{equation}
	\label{eq:A7:b}
	z_2 = \frac{e^{-\beta \epsilon_{\mathrm b}}}{ e^{\beta g} + 2
	e^{-\beta g}}.
\end{equation}
\end{subequations}
The first factor in Eq.~(\ref{eq:A7:a}) is the contribution of the
random walk of the free strand  when the DNA is in the unzipped phase
and the second factor is the contribution of the force on the pulled
strand. The phase boundary separating the zipped and the unzipped
phases is given by $z_2 = z_4$, which reads as
\begin{equation}
	\label{eq:A8}
	g(T) = \frac{k_{\mathrm B}T}{2} \ln \left( \frac{ 2 e^{-\beta
	\epsilon_{\mathrm b}} - 2} { 1 - 2 e^{-\beta \epsilon_{\mathrm b}}}
	\right).
\end{equation}
The zero force limit (i.e. $g \rightarrow 0$) of Eq.~(\ref{eq:A8}) gives
the thermal melting temperature of the dsDNA. This gives 
\begin{equation}
	\label{eq:A8a}
	T_{\mathrm m} = \frac{\epsilon_{\mathrm b}} {\ln \ 4/3 }.
\end{equation}
In the opposite limit (i.e. $g \rightarrow \infty$), we obtain the
minimum temperature 
\begin{equation}
	\label{eq:A8b}
	T_{\mathrm u} = \frac{\epsilon_{\mathrm b}} { \ln 2},
\end{equation}
above which the dsDNA can be unzipped by pulling a single strand.

By using the same analysis given above, we can also approximate the
phase boundaries between the zipped and the unzipped phases in the
presence of an attractive surface with energy $-\epsilon_{\mathrm w}$
near DNA. By maintaining the structure of the singularity $z_4$, the
singularity which contribute to the partition function of phase \phIII
~in the thermodynamic limit can be approximated as
\begin{equation}
	\label{eq:A9}
	z_{3} = \left( \sqrt{ e^{-\beta \epsilon_{\mathrm w}} - e^{-2
	\beta \epsilon_{\mathrm w}} } \right) \left( \frac{1}{ 2 \cosh \beta
	g} \right).
\end{equation}
In Eq.~(\ref{eq:A9}), the first factor is the contribution of the
adsorbed free strand on the surface and the second factor is same as
in Eq.~(\ref{eq:A7:a}). The phase boundary separating phase \phII ~with
phase \phIII ~is given by 
\begin{equation}
	\label{eq:A10}
	g(T) = \frac{k_{\mathrm B} T}{2} \ \ln \left( \frac{ 2 \sqrt{
	e^{-\beta \epsilon_{\mathrm w}} - e^{-2 \beta \epsilon_{\mathrm w}}
	} - e^{-\beta \epsilon_{\mathrm b}} } { e^{-\beta \epsilon_{\mathrm
	b}} - \sqrt{ e^{-\beta \epsilon_{\mathrm w}} - e^{-2 \beta
	\epsilon_{\mathrm w}}}} \right),
\end{equation}
which is obtained by equating $z_{4}$ with $z_2$. This phase boundary is
shown by dotted lines in Fig.~\ref{fig:phdia}. The approximation done in
Eq.~(\ref{eq:A9}) is valid only for smaller values of $\epsilon_{\mathrm
w}$ and breaks as $\epsilon_{\mathrm w}$ becomes comparable to
$\epsilon_{\mathrm b}$ as evident from Fig.~\ref{fig:phdia}(b) and (c).
However, the $g \rightarrow \infty$ limit of Eq.~(\ref{eq:A10}) gives
the $T_{\mathrm u}$ and $\epsilon_{\mathrm w}$ dependence as 
\begin{equation}
	\label{eq:A11}
	\epsilon_{\mathrm w} =  k_{\mathrm B} T_{\mathrm u} \ \ln \left[
	\frac{2} {1 - \sqrt{ 1 - 4 \exp \left(-2 \epsilon_{\mathrm b} /
	k_{\mathrm B} T_{\mathrm u} \right) } }  \right],
\end{equation}
from which one obtains $T_{\mathrm u} \rightarrow \epsilon_{\mathrm b}/
\ln 2$ as $\epsilon_{\mathrm w} \rightarrow 1$ and $T_{\mathrm u}
\rightarrow 0$ as $\epsilon_{\mathrm w} \rightarrow 2$. 

\renewcommand{\theequation}{B-\arabic{equation}}
\setcounter{equation}{0}  

\section*{APPENDIX B: DETAILS OF MOTE CARLO SIMULATIONS } 

In this appendix we give the details of Monte Carlo simulations used in
obtaining the force-distance isotherms in both $1+1$ and $2+1$
dimensions. We model the bases of the DNA by beads and the two adjacent
beads on a strand, are connected by a rigid rod of unit length which
stays on edges of the square or the cubic lattice in $1+1$ and
$2+1$ dimensions respectively. If the beads of the two strands are unit
distance apart, there is a binding between them. At the diagonal of the
square and the cube there is an impenetrable surface. One end of both
the chains are anchored at the origin and a pulling force $g$ is applied
on the bead at the free end of the pulled chain. We consider a single
bead flip dynamics. In $1+1$ dimensions, the $j{\rm th}$ bead in the
interior of the strand, e.g. the free strand, located at a distance
$x_{1_j}$ from the surface (state $\mu$), is flipped to
$x_{1_j}^{\prime} = x_{1_j} \pm 2$ (state $\nu$), provided $x_{1_{j+1}}
- x_{1_j}^{\prime} = x_{1_j}^{\prime} - x_{1_{j-1}} = 1$. This
constraint ensures that the chain does not break while doing the
dynamics. The move is accepted (or rejected) according to the rule
\begin{equation} \label{rule}
	P(\mu \rightarrow \nu) = 
	\begin{cases} 
		e^{ - \beta (E_{\nu} - E_{\mu}) } & \text{if} \quad E_{\nu} -
		E_{\mu} > 0 \cr 
		1 & \text{otherwise}.
	\end{cases}
\end{equation}
The same thing has to be repeated for the pulled strand also. For the
end bead we have to take care of the energy contribution
$g(x_{2_N}^{\prime} - x_{2_N})$ due to the force $g$. In $2+1$
dimensions, the bead can be flipped in three positions without breaking
the chain. Apart from the connectivity constraint, the excluded volume
effects, between the strands and between the free strand and the
surface, have to be taken care of. This means the following moves have
to be rejected while doing the dynamics (i) the free chain crossing the
pulled chain, (ii) the free chain crossing the surface on which it is
adsorbed, and (iii) the pulled chain crossing the first chain. It is
possible for the DNA to reach any state from any other state using the
above moves. The $2N$ such flips, $N$ for each strand, constitutes one
Monte Carlo step (MCS) per bead. At each value of the force, where the
simulation is performed, we allow the system to equilibrate by repeating
the above procedure for $10^6$ MCS and start measurements only after it.
Between any two measurements we run the procedure, without doing
measurements, for $10^3$ MCS to avoid correlations between two
measurements.

We have performed $K=7$ simulations at $\beta = 2/3$ at various $g$ as
indicated in Fig.~\ref{fig:xg}(b). The expectation
value of $\langle x_{1,2} \rangle$ is obtained by averaging over $10^6$
MCS. Other than the averaged values, we have also obtained histograms,
$h_k(E, x_1, x_2) (k=1, \ldots K)$ of the energy $E$ (binding from the
surface plus base pairing but not the contribution from the force), the
positions of the end monomers for the free and the pulled strands
represented by $x_1$ and $x_2$ respectively. The $k{\rm th}$ histogram
is collected at force $g_k$ with $n_k$ number of states. By using the
multiple histogram technique~\cite{ferren89}, the average distance (from
the surface) of the last monomers of both strands, $\langle x_{1,2}
\rangle$ at force $g$ is given by 
\begin{widetext}
\begin{equation}\label{eq:mhistQ}
	\langle x_{1,2} (\beta, g) \rangle = \frac{1}{Z(\beta, g)} \sum_{E, x_1,
	x_2} x_{1,2} \ e^{-\beta( E - g x_2)} \frac{ \sum_{j=1}^{K} h_j(E,
	x_1, x_2)} { \sum_{j=1}^{K} n_j Z_j^{-1} e^{ - \beta (E - g_j x_2)
	} },
\end{equation}
\end{widetext}
where
\begin{equation}\label{eq:partbg}
	Z(\beta, g) = \sum_{E, x_1, x_2} e^{ -\beta ( E - g x_2 ) } \frac{
	\sum_{j=1}^{K} h_{j}(E, x_1, x_2) }{ \sum_{j=1}^{K} n_j Z_j^{-1} e^{
	- \beta (E - g_j x_2)} },
\end{equation}
is an approximate partition function at $g$, estimated
using $Z_k$, i.e., the partition functions at $g_k$.
The partition functions $Z_k$ are obtained self-consistently from the
following equation
\begin{equation}\label{eq:part}
	Z_{k} = \sum_{E, x_1, x_2} e^{ -\beta ( E - g_k x_2) } \frac{
	\sum_{j=1}^{K} h_{j}(E, x_1, x_2) }{ \sum_{j=1}^{K} n_j Z_j^{-1} e^{
	- \beta (E - g_j x_2)} }.
\end{equation}
We iterate the above equation with starting values $Z_k = 1$ for all
$k$. The convergence is monitored by estimating the amount of change
after each iteration. We gauged it by calculating the quantity
\begin{equation}\label{appC:eq:gauge}
	\Delta^2 = \sum_k \left[ \frac{ Z_k^{(m)} - Z_k^{(m-1)} } {
	Z_k^{(m)} } \right]^2,
\end{equation}
where $Z_k^{(m)}$ is the value of $Z_k$ at the $m{\rm th}$ iteration.
When this quantity falls below some predefined quantity $\epsilon^2$,
the convergence is achieved. We take $\epsilon^2 = 10^{-14}$.
The force-distance isotherms obtained by this procedure in $1+1$
dimensions is shown in Fig.~\ref{fig:xg}(b).


\begin{thebibliography}{10}
\bibitem{cep}
	A critical end point is a critical point where a line of
	second-order transition terminates at a line of first-order
	transition.
\bibitem{tp}
	A triple point is the intersection on a phase diagram where three
	phases coexist in equilibrium.
\end{thebibliography}
\end{document}